\documentclass[reprint,aps,prx,floatfix]{revtex4-2}
\usepackage{newtxtext, newtxmath}  % Times font to match PRX.
\usepackage{graphicx, braket, siunitx, mathtools, dcolumn, hyperref}

\DeclareMathOperator\Tr{Tr}
\DeclareSIUnit[number-unit-product=]\percent{\char`\%}

\begin{document}

\title{Certifying Multilevel Coherence in the Motional State of a Trapped Ion}

\author{Ollie Corfield}
\thanks{These two authors contributed equally.}
\affiliation{Blackett Laboratory, Imperial College London, London SW7 2AZ, United Kingdom}

\author{Jake Lishman}
\thanks{These two authors contributed equally.}
\affiliation{Blackett Laboratory, Imperial College London, London SW7 2AZ, United Kingdom}

\author{Chungsun Lee}
\affiliation{Blackett Laboratory, Imperial College London, London SW7 2AZ, United Kingdom}

\author{Jacopo Mosca Toba}
\affiliation{Blackett Laboratory, Imperial College London, London SW7 2AZ, United Kingdom}

\author{George Porter}
\affiliation{Blackett Laboratory, Imperial College London, London SW7 2AZ, United Kingdom}

\author{Johannes M. Heinrich}
\affiliation{Blackett Laboratory, Imperial College London, London SW7 2AZ, United Kingdom}

\author{Simon C. Webster}
\affiliation{Blackett Laboratory, Imperial College London, London SW7 2AZ, United Kingdom}

\author{Florian Mintert}
\affiliation{Blackett Laboratory, Imperial College London, London SW7 2AZ, United Kingdom}

\author{Richard C. Thompson}
\affiliation{Blackett Laboratory, Imperial College London, London SW7 2AZ, United Kingdom}

\date{\today}

\begin{abstract}
Quantum coherence is the foundation of almost all departures from classical physics, and is exhibited when a quantum system is in a superposition of different basis states.
Here the coherent superposition of three motional Fock states of a single trapped ion is experimentally certified, with
a procedure that does not produce false positives.
As the motional state cannot be directly interrogated, our scheme uses an interference pattern generated by projective measurement of the coupled qubit state.
The minimum number of coherently superposed states is inferred from a series of threshold values based on analysis of the interference pattern.
This demonstrates that high-level coherence can be verified and investigated with simple, nonideal control methods well-suited to noisy intermediate-scale quantum devices.
\end{abstract}

\maketitle

\section{Introduction}
\label{sec:introduction}

A defining feature of quantum mechanics is the ability for a system to be in a coherent superposition of any set of conceivable states.
Coherence between states of a natural basis underpins almost all deviation between the predictions of quantum and classical mechanics, however theoretical efforts to rigorously define and it are relatively recent~\cite{Oi2006,Girolami2014,Streltsov2017}.
Similar to entanglement, quantum coherence is recognised as a resource~\cite{Levi2014,Baumgratz2014,Ringbauer2018} that may be expended to realise desirable outcomes, such as improving the probability of success in quantum information algorithms~\cite{Hillery2016,Shi2017} and phase-estimation metrology applications~\cite{Castellini2019}, the extraction of thermodynamic work~\cite{Korzekwa2016}, or the creation of nonequilibrium entropy~\cite{Santos2019}.

Verification of coherence in classical systems is not limited by the inherent restriction of measurements in quantum systems, and the rigorous verification of quantum coherence requires mathematical tools similar to those developed in entanglement theory.
Previous work on quantum coherence metrics has typically focused on producing a quantity determined by complete reconstruction of the density matrix~\cite{Baumgratz2014,Winter2016,Yuan2015}.
In practice, the poor scaling of the number of precise measurements required to perform such state tomography renders it an unattractive prospect beyond small-dimensioned systems.
These measures also explore little of the concept of higher-order coherence~\cite{Levi2014,Sperling2015}, an analogue of multipartite entanglement, where the quantity of interest is the number of basis states that contribute to a quantum superposition.
This multilevel coherence is known to be its own resource within certain quantum information processing operations~\cite{Ringbauer2018}, and is critical to the deeper understanding of quantum transport in regimes of partially coherent dynamics, both in solid-state physics~\cite{Levi2014} and transport processes on the boundary between coherent and dissipative dynamics~\cite{Lambert2013,Cao2020}.

Early efforts to quantify multilevel coherence required the ability to measure in bases that themselves contained some level of coherence~\cite{Levi2014,VonPrillwitz2015}.
As coherence is defined purely with respect to a basis, usually the only one available for measurements, achieving these schemes experimentally necessitated an extra coherent step.
The tested state would undergo several assumed-coherent operations to map the ideal coherent measurement basis onto the actual, typically incoherent, basis of the measurement.
Information derived from such experiments would then rely on these operations having been performed accurately.
However, these same types of operations also form the state creation, which is what is being tested; the operations must therefore be both trusted and tested simultaneously.
More recent schemes have sought to avoid these problems by ensuring that their coherence metrics provide a strict lower bound on the amount of higher-order coherence in a state, no matter how successfully intermediate operations were implemented~\cite{Ringbauer2018}.
Of particular interest is the ability to certify this coherence based only on an interference pattern~\cite{Dive2020}, effectively enabling verification of multilevel coherence based on a projective measurement onto only one state.
These schemes enable insight into coherence in systems inaccessible by measurement, without risking false positives from imperfect mapping operations.
This is of particular interest for noisy intermediate-scale quantum devices, where verification of quantum properties is a laborious task.

Trapped ions are well established as media for quantum simulation~\cite{Blatt2012,Lanyon2011}, and high-precision metrology via quantum logic spectroscopy~\cite{Schmidt2005,Rosenband2007}.
Additionally they are one of the leading candidates for a full-scale quantum computer~\cite{Monroe2014,Bruzewicz2019,Ballance2016,Gaebler2016}.
Quantum information is encoded in trapped ions in two different forms.
The coupled motion can be cooled into the quantum regime, and used as a means to drive entangling interactions between internal qubit states.
Decoherence processes of the motional modes are among the dominant effects reducing quantum logic fidelities, making their classification and understanding imperative, however only the qubit states can be measured directly.
While it is possible to fully reconstruct the density matrix and Wigner function of an arbitrary motional state~\cite{Leibfried1996}, this requires a large number of measurements.

In this work, we experimentally certify the existence of multiple superposition elements in the motional state of a single trapped ion, using an interference pattern method derived from, and extending, ref.~\cite{Dive2020}.
The only available projective measurement can distinguish the state of the coupled internal-state qubit, but not the different motional states, making it a more general operation than originally considered.
We extend the theory by showing that the same method is valid for arbitrary measurement operators, and experimentally show that coherence can be created and verified in the Fock basis of the motion of trapped ions using only the simplest operations on a noisy device.

\section{Implementation}
\label{sec:implementation}

The system Hamiltonian of a single two-level ion in a harmonic trap, considering only a single motional mode with annihilation operator $\hat a$, is
\begin{equation}\label{eq:system-hamiltonian}
\hat{\mathcal H}/h = \frac12\nu_{eg}\hat\sigma_z + \nu_m{\hat a}^\dagger\hat a.
\end{equation}
Here, $\nu_{eg}$ is the frequency separation of two qubit states $\ket g$ and $\ket e$, and $\nu_m$ the trap frequency, which gives the uniform separation of oscillator Fock states $\ket n$ representing physical quanta of motion in the system.
In principle the motional states can contain arbitrarily high levels of coherence, making them ideal candidates for investigation, though hampered by the lack of direct measurement available.
The experimental aim is to create arbitrary superpositions of different Fock states in the motion, and certify their multilevel coherence properties using only simple operations and measurements of the qubit.

% Trap
The qubit is mapped onto the $\text{S}_{1/2,m_j=1/2} \leftrightarrow \text{D}_{5/2,m_j=1/2}$ quadrupole $\pi$ transition of a $^{40}\text{Ca}^+$ ion trapped in a linear rf trap.
The motional states are the quantised levels of the axial mode, with a frequency separation $\nu_m\approx\SI{1.1}{\mega\Hz}$.
At the start of each experimental run, the ion is prepared in the $\ket{g,0}$ state by Doppler and sideband cooling, with a \SI{98(2)}{\percent} probability of success.
The qubit state is read out by a fluorescence measurement with electron shelving, giving a fidelity reliably above \SI{99}{\percent}.

% Sideband Pulses
Coherent population transfer between the qubit states is driven by a sub-\SI{1}{\kilo\Hz}-linewidth diode laser tuned close to the qubit frequency $\nu_{eg}\approx\SI{411}{\tera\Hz}$.
The motion-preserving carrier transition of the qubit has a coupling strength characterised by the Rabi frequency $\Omega$.
Motion-altering transitions, achieved by detuning the drive by a multiple of the motional frequency, are additionally dependent on the Lamb--Dicke parameter $\eta=k\sqrt{\hbar/(4\pi m\nu_m)} \approx 0.09$, where $m$ is the ion mass and $k$ is the laser wavevector.
This is inside the Lamb--Dicke regime of weak qubit--motion coupling, where the only three available processes are the carrier, and the red (blue) sideband that removes (adds) a phonon of motion when exciting the qubit.
These two sidebands have coupling frequencies of $\eta\Omega\sqrt n$ when coupling the motional states $\ket{n-1}$ and $\ket{n}$, and so drive nonperiodic evolution.
As they are weaker transitions, the two sidebands have significant undesired off-resonant effects from the presence of the nearby carrier, which reduce the fidelity of their operations.
The AC Stark effect is mitigated by applying compensation pulses far-off-resonantly on the opposite side of the carrier, but there remain small-amplitude oscillations between the qubit states, which cannot reliably be nulled on completion of the pulse.

\begin{figure}
    \includegraphics{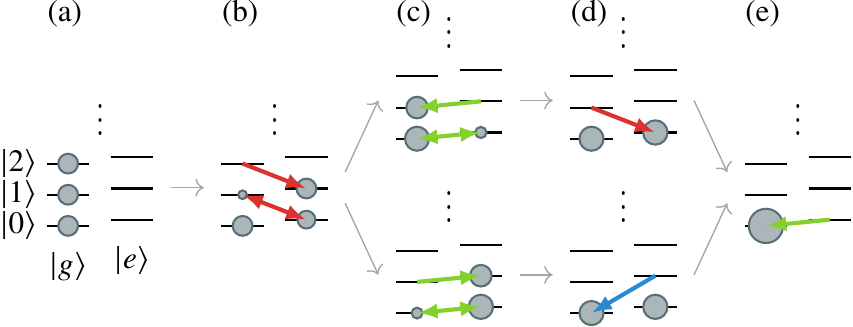}
    \caption{\label{fig:superposition-preparation2}%
        The algorithm used to produce arbitrary motional superpositions~\cite{Gardiner1997}, illustrated creating $\bigl(\ket{g,0} + \ket{g,1} + \ket{g,2}\bigr)/\sqrt3$.
        Grey circles represent superposition state occupation with size proportional to population, while green, red and blue arrows represent carrier, red and blue sideband transitions respectively.
        (a) Consider the system beginning in the target state.
        (b) A red-sideband pulse is applied to move all of the highest-occupied motional state $\ket2$ into the state with one quantum of motion fewer; this affects all other states by different amounts that must be tracked.
        (c) The now-highest-occupied motional state $\ket1$ has population in both qubit states, so the carrier is used to combine both into either $\ket{g,1}$ or $\ket{e,1}$.
        (d) Depending on the previous pulse, the red or blue sideband is used to reduce the highest motional state again, combining populations in $\ket{g,1}$ and $\ket{e,0}$ (red), or $\ket{e,1}$ and $\ket{g,0}$ (blue), into the motional ground state.
        (e) A final carrier combines the population into $\ket{g,0}$.
        The desired forwards operation is simply the adjoint of the derived sequence.
    }
\end{figure}

\begin{figure*}
    \includegraphics[width=\linewidth]{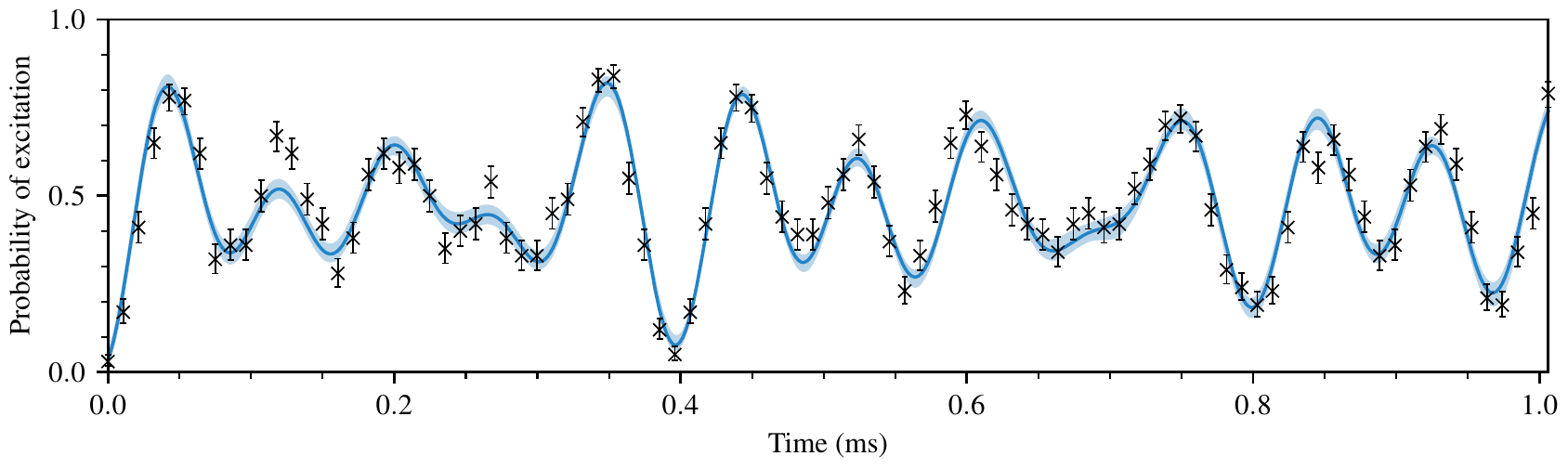}
    \caption{\label{fig:bsb}%
    Time evolution of state $\bigl(\ket{g,0} + \ket{g,1} + \ket{g,2}\bigr)/\sqrt3$, while driving the blue-sideband transition. 
    The data points (black crosses) are the measured excitation probabilities with Wilson binomial 1-$\sigma$ confidence bounds indicated by error bars.
    The best fit (darker blue line) was found by maximum-likelihood estimation, and the 95\% confidence region (lighter blue shaded region) by bootstrapping the measured data \num{14000} times.
    The fit was made over the sideband Rabi frequency, sideband detuning, motional dephasing rate, and a reduced density matrix including correlations only between directly coupled elements.
    The Rabi frequency coupling the $\ket{g,0}$ state was estimated at \SI{6.94(3)}{\kilo\hertz}, with oscillation from $\ket{g,n}$ increased by a factor of $\sqrt{n+1}$.
    The populations in $\ket{g,0}$, $\ket{g,1}$ and $\ket{g,2}$ were \SI{33(2)}{\percent}, \SI{30(2)}{\percent} and \SI{33(2)}{\percent}, respectively, with \SI{4.7(14)}{\percent} outside the desired basis elements, where the errors denote the 1-$\sigma$ confidence region from bootstrapping.
    These values all have significant negative covariance, as expected.
    All appreciable undesired population was in the $\ket e$ excited qubit state; the motional state $\ket3$ was included in the fit, but found to have a population consistent with zero with a standard error of \num{9e-3} percentage points.
}
\end{figure*}

% State preparation
The motional superpositions are created using an extension to previous work that can synthesise any joint qubit--motion state in a single trapped ion~\cite{Gardiner1997,Ben-Kish2003}.
This takes advantage of how the $\ket{g,0}$ ($\ket{e,0}$) state is unaffected by the red (blue) sideband, and is illustrated in fig.~\ref{fig:superposition-preparation2}.
Considering the operation in reverse and starting from the target state, the motional excitation is gradually stepped downwards either directly by a sideband pulse, or by combining two populations so that the next step may continue the descent.
At the end, the adjoint of the sequence is taken to produce the forwards operation.
Whenever the carrier is used, the population can be combined into either $\ket g$ or $\ket e$ as in fig.~\ref{fig:superposition-preparation2}c, and at every step there are an infinite number of pulse lengths that will achieve the desired population transfer.
It may be advantageous to choose a longer early pulse, or prefer combining population in one qubit state over the other, in order to realise time gains later.
After an initial candidate solution is found by arbitrarily resolving the choices, the minimal-time solution can be found by recursively trying every path, and pruning solutions that become too long.

After creation of a state, 
the motional populations can be measured by a Rabi-type experiment; the blue sideband is applied for a varying amount of time, followed by a projective measurement on the qubit.
This produces an oscillatory pattern, shown in fig.~\ref{fig:bsb} for the state $(\ket{g,0}+\ket{g,1}+\ket{g,2})/\sqrt{3}$, containing components at frequencies proportional to $\sqrt{n+1}$ for small numbers of phonons $n$, whose amplitude is equal to the population in the respective motional state.
This does not give any information about the coherence properties of the state however.

\section{Coherence Certifier}
\label{sec:certifier}

The coherence of a system is defined with respect to a particular basis.
The eigenstates of the system Hamiltonian are a natural choice, typically having a clear physical interpretation, and are often the only readily accessible measurement basis.
A pure state $\ket\psi = \sum_j \zeta_j\ket{j}$ is commonly called $k$-coherent in a basis $\{\ket j\}$ if there are at least $k$ nonzero coefficients $\zeta_j$.
A mixed state $\rho$ is then $k$-coherent if there is no pure-state decomposition $\rho = \sum_i q_i \ket{\psi_i}\!\bra{\psi_i}$ without at least one of the $\ket{\psi_i}$ being $k$-coherent and having nonzero probability $q_i$.
This is similar to equivalent definitions used when classifying entanglement~\cite{Szalay2015}, and produces a discrete hierarchy for coherence.

A coherence metric is a single number, measurable by a simple experiment, that imparts some information on the coherence level of the given state.
For a certifier, any concrete result must be unimpeachable.
The responsibility is principally to ensure that a false positive can never occur, and only within this to maximise the information returned.
A statistic of an experiment may certify coherence if, for each level of coherence, there is a theoretically maximal value achievable by any state at this level, and these values strictly increase with respect to the level.
The maximal values then form a series of thresholds: measuring a value greater than is possible for a $k$-coherent state unambiguously implies that the state was at least $(k{+}1)$-coherent.
We call the certifier robust if, in addition, there is no possible error in its measurement that could cause it to exceed a threshold it would not exceed with perfect measurement.
A robust coherence certifier may not always be able to certify the actual level of coherence, but it will always provide a lower bound.

\subsection{Interference-Pattern Methods}

Interference patterns are well known as a tool for certifying quantum coherence between two states.
Any evidence of oscillatory behaviour in the common Ramsey experiment on a quantum system is sufficient to verify 2-coherence.
It is desirable to continue in this vein; interference patterns are a function of only one control parameter and are produced by simple projective measurements onto one basis state, so are easily experimentally achievable.

The particular interference patterns considered here are higher-order extensions of the Ramsey experiment.
An input state $\rho$ to be tested for coherence undergoes evolution through a Hamiltonian with a single controllable parameter, before a predetermined pulse sequence is applied to map the coherent basis of interest back to the measurement basis.
The dynamics of this measurement with respect to the control parameter determine the interference pattern.
Here, as the system under consideration is a harmonic oscillator, the dynamics $\hat{\mathcal U}_{\text f}$ are free evolution under eq.~\eqref{eq:system-hamiltonian}, and the control parameter is the phase $\phi$ of the period of oscillation.
Explicitly, the interference pattern is
\begin{equation}
\label{eq:interference-pattern-rank1-projector}
p(\phi) = \bra\chi\hat{\mathcal U}_{\text m} \hat{\mathcal U}_{\text f}(\phi) \rho\, \hat{\mathcal U}_{\text f}^\dagger(\phi) \hat{\mathcal U}_{\text m}^\dagger\ket\chi
\end{equation}
where $\hat{\mathcal U}_{\text m}$ is the operation mapping the coherent basis to the measurement basis, and $\ket\chi$ is the basis state whose population is measured.
In the special case of the two-state Ramsey sequence, an optimal measurement mapping is the adjoint of the operation used to create the initial superposition.

In degenerate systems, it is more appropriate to use an easily realised evolution Hamiltonian that breaks the degeneracy.
For example, in a system of $d$ qubits with the coherence defined over the product of $z$-basis eigenstates, the Hamiltonian $\hat{\mathcal R}_1(\phi)\hat{\mathcal R}_2(2\phi)\dotsm\hat{\mathcal R}_d(2^{d-1}\phi)$, where $\hat{\mathcal R}_k(\phi)$ is a $z$-rotation by $\phi$ of qubit $k$, can be realised with only single-qubit operations.
Under this evolution each basis element, such as $\ket{011}$ or $\ket{101}$, would gain a phase equal to $\phi$ multiplied by the binary value of its label, permitting a full-contrast interference pattern.

The certifier $C$ used here is a function of normalised moments
\begin{equation}
M_n = \frac1{2\pi}\int_0^{2\pi}{p(\phi)}^nd\phi,
\end{equation}
of this interference pattern.
No single moment is suitable taken alone, however the ratio
\begin{equation}
C = \frac{M_3}{M_1^2},
\end{equation}
satisfies the requisites~\cite{Dive2020}.
It requires the pattern to be evaluated at only a few different free-evolution phases for good statistics, as it relies only on low-order moments.
Obtaining a value of $C$ greater than $1$ requires 2-coherence, greater than $5/4=1.25$ requires 3-coherence, and greater than $179/96\approx1.86$ is necessary to certify 4-coherence in a Hilbert space of arbitrary dimension~\cite{Dive2020}.

\subsection{General Measurements}

Although the coherence metric is resilient to imperfect operations during the measurement-mapping procedure, the analysis of ref.~\cite{Dive2020} is valid only when the actual measurement is a projection onto a single basis state.
In order to be used for coupled systems, it must first be shown that the same hierarchical structure of the threshold values applies for measurements beyond simple projections.
This can be demonstrated by showing that the thresholds for determining certain levels of coherence remain the same, regardless of the type of measurement.

The interference pattern in eq.~\eqref{eq:interference-pattern-rank1-projector} is generalised to 
\begin{equation}
\label{eq:interference-pattern-general}
p(\phi) = \Tr\bigl[\hat{\mathcal A}\hat{\mathcal U}_{\text m} \hat{\mathcal U}_{\text f}(\phi) \rho\, \hat{\mathcal U}_{\text f}^\dagger(\phi) \hat{\mathcal U}_{\text m}^\dagger\bigr],
\end{equation}
where $\hat{\mathcal A}$ is an element of a positive operator-valued measure (POVM).
Projections onto a single basis state are rank-1 matrices with $\hat{\mathcal A} = \ket\chi\!\bra\chi$, for a measurement-basis state $\ket\chi$.
The only measurement available in the ion trap is a projective measurement on the qubit state only, such as $\hat{\mathcal A} = \ket g\!\bra g\otimes\hat{\mathbb I}_{\text{mot}}$, where $\hat{\mathbb I}_{\text{mot}}$ is the identity operator on the motional space.
This is a type of higher-rank projective measurement, but the proof is applicable to general measurements.

As in ref.~\cite{Dive2020}, the analysis is performed in terms of a harmonic oscillator with eigenstates $\{\ket n\}$.
Any periodic free-evolution can be modelled as such by inserting non-interacting states with the otherwise-absent intermediate energy levels.
This includes Hilbert spaces with tensor-product structure, as in trapped ions, by relabelling the states with a single index $n$.
Using the decompositions $\rho = \sum_{n,m}\rho_{nm}\ket n\!\bra m$ and $\hat{\mathcal A} = \sum_{n,m}A_{nm}\ket n\!\bra m$, the interference pattern can be written as
\begin{equation}
p(\phi) = \sum_n \rho_{nn}A_{nn} + 2\sum_{n>m}\lvert\rho_{mn}A_{nm}\rvert\cos\bigl((n-m)\phi+\theta_{nm}\bigr),
\end{equation}
where $\theta_{nm}$ is the complex phase of $\rho_{mn}A_{nm}$.

In this form, the value of the first moment of the interference pattern $M_1$ is the sum of the $\phi$-independent terms.
The only nonzero terms of the moment $M_3$ are each proportional to $\cos(\theta_{n_1,m_1} \pm \theta_{n_2,m_2} \pm \dotsb)$, so the certifier $C$ will reach a maximal value when all the $\theta$ are zero, and $\rho$ and $\hat{\mathcal A}$ can be taken as real-symmetric matrices without loss of generality in determining threshold values.

The maximum achievable value of $C$ for a 2-coherent state under these general measurements can be verified analytically.
To be less than 3-coherent, $\rho$ may have at most two nonzero diagonal elements in the coherence basis, and one upper-triangular off-diagonal entry.
The value of $C$ is not dependent on particular energy levels, so for simplicity these are labelled $0$ and $1$.
The convexity of $C$ is unaffected by the general measurement, so it is sufficient to consider only pure states.
The state can be parametrised by a real value $0\le x\le1$ as $\sqrt{x}\ket0 + \sqrt{1-x}\ket1$.
The value of the certifier is then
\begin{equation}
\label{eq:minimisation-target}
C = xA_{00} + (1 - x)A_{11} + \frac{6x(1-x)A_{01}^2}{xA_{00} + (1-x)A_{11}}.
\end{equation}
In order for the measurement operator to be a valid POVM value, the two diagonal elements must individually be between 0 and 1, and the off-diagonal element must satisfy
\begin{equation}
\label{eq:povm-off-diagonal-inequality}
A_{01} \le \min\Bigl\{A_{00}A_{11},\,(1-A_{00})(1-A_{11})\Bigr\}.
\end{equation}
It is clear that if $x$, $A_{00}$ or $A_{11}$ are either 0 or 1 exactly, this is equivalent to the incoherent case, and the maximum value of $C$ is 1.
Further, $C$ is always maximised by maximising the magnitude of the off-diagonal element $A_{01}$, and so only the equality in eq.~\eqref{eq:povm-off-diagonal-inequality} need be considered.
The two branches of the minimum correspond to using $\hat{\mathcal A}$ or $1-\hat{\mathcal A}$ as the measurement, so without loss of generality it is possible to consider only $A_{01} = A_{00}A_{11}$ and $A_{00} + A_{11} \le 1$.

The maximal value of $C$ can be found from eq.~\eqref{eq:minimisation-target}, using the method of Lagrangian multipliers with the constraints $0 < \{x,\,A_{00},\,A_{11}\} < 1$ and $A_{00} + A_{11} \le 1$.
Only the latter bound can be tight, so the Lagrangian can be written as
\begin{equation}\mathcal L = C - \lambda(A_{00} + A_{11} - 1),\end{equation}
with $\lambda \ge 0$.
The derivative with respect to $A_{00}$ is
\begin{equation}
\frac{\partial\mathcal L}{\partial A_{00}} =
    x\frac{x^2 A_{00}^2 + 2x(1-x)A_{00}A_{11} + 7{(1-x)}^2A_{11}^2}{{\bigl[xA_{00} + (1-x)A_{11}\bigr]}^2} - \lambda,
\end{equation}
and the derivative with respect to $A_{11}$ is the same under the transformations $x\to1-x$ and $A_{00}\leftrightarrow A_{11}$.
Under the constraints, the fraction is strictly greater than zero, so when also satisfying the complementary slackness condition $\lambda(A_{00} + A_{11} - 1)=0$, the stationary points of the Lagrangian all have $A_{11} = 1 - A_{00}$.
With this, the optimal measurement operator can be written in the $\{\ket0,\,\ket1\}$ basis as
\begin{equation}
\hat{\mathcal A} = \begin{pmatrix}
A_{00} & \sqrt{A_{00}(1-A_{00})} \\
\sqrt{A_{00}(1-A_{00})} & 1 - A_{00}
\end{pmatrix},
\end{equation}
or similarly in the full coherence basis with padding zeros in all other positions.
This is exactly the form of a rank-1 projective measurement on the state $\sqrt{A_{00}}\ket0 + \sqrt{1-A_{00}}\ket1$, and consequently the maximal value of $C$ obtainable with a general measurement on a 2-coherent state remains $5/4$, using the proofs given in ref.~\cite{Dive2020}.
It is therefore sufficient to observe a value of $C$ greater than $5/4$ to certify that the underlying state is at least 3-coherent, no matter the type of measurement used.
Numerical optimisations are strongly suggestive that the thresholds remain the same for all orders of multilevel coherence.
Details of how these were performed may be found in Appendix A.

\subsection{Measurement-Mapping Sequences}
\label{sec:implementation-certification}

\begin{table*}%
    \newcolumntype{x}{D..{1.2}}%
    \begin{ruledtabular}\begin{tabular}{l@{\hskip 1.5em}xxxxc@{\hskip 1.5em}xxxxx}
    & \multicolumn{4}c{State creation} && \multicolumn{5}c{Measurement mapping}\\
        \rule{0pt}{2.5ex}Transition & \multicolumn1c{carrier} & \multicolumn1c{red} & \multicolumn1c{carrier} & \multicolumn1c{red} && \multicolumn1c{red} & \multicolumn1c{carrier} & \multicolumn1c{red} & \multicolumn1c{carrier} & \multicolumn1c{red}\\
        Pulse length          &  0.50 &  0.70 &  0.73 &  0.71 &&  0.71 &  0.50 &  1.42 &  1.59 &  0.72\\
        Phase offset ${}/\pi$ &  0    & -0.50 &  1.00 &  0.50 &&  0    &  0.71 & -0.29 &  0.10 & -0.51
    \end{tabular}\end{ruledtabular}%
    \caption{\label{tab:pulses}%
        Pulse sequence for creation and measurement mapping of target state $\bigl(\ket{g,0} + \ket{g,1} + \ket{g,2}\bigr)/\sqrt3$.
        Only carrier and red sideband transitions are used.
        The pulse length is scaled relative to the oscillation frequency of the coupled pair that includes the motional $\ket0$ state, so that a value of $1$ is the time taken to exchange $\ket{g,0}\leftrightarrow\ket{e,0}$ on the carrier and $\ket{g,1}\leftrightarrow\ket{e,0}$ on the red sideband.
        The given phase is applied as an offset, so that the set laser phase at the beginning of a pulse is offset relative to where it would have been had it oscillated freely at its transition frequency since the beginning of the experiment.
        The interference pattern is constructed by adding a varying phase offset on the red-sideband pulses during the measurement mapping.
    }
\end{table*}

With the impossibility of a false positive proved, attention must now be given to maximising the amount of conclusive information that is returned.
The only available measurement in the ion-trap system can distinguish the two qubit states, but not the motional basis states by which the superpositions are defined.
This is a type of high-rank projective measurement, which were shown to have worse performance than rank-1 projectors in the previous section.
The qubit measurement can be converted into an effective rank-1 projective measurement by using a measurement-mapping operation that maps the expected state onto one qubit state, and the rest of the space spanned by the free evolution of the expected state onto the other qubit state.

The standard two-state Ramsey experiment achieves this for a target state of $(\ket{g,0} + \ket{g,1}) / \sqrt2$, using a measurement mapping that is the inverse of the simplest creation protocol.
This simple inversion is insufficient in general, however, as the adjoint of the creation operation does not necessarily map states that are orthogonal to the target into the other qubit state.

For a target state with $n>2$ populated motional superposition elements, the mapping will need to map several different motional states to one qubit state.
Only states that are orthogonal to the target and reachable by free evolution need be considered.
As an example, the measurement mapping $\hat{\mathcal U}_{\text m}$ for a target state $(\ket{g,0} + \ket{g,1} + \ket{g,2})/\sqrt3$ should satisfy
\begin{equation}\label{eq:measurement-mapping}\begin{aligned}
\hat{\mathcal U}_{\text m} \bigl(\ket{g,0} + \ket{g,1} + \ket{g,2}\bigr) &\propto \ket{e,\lambda_1},\\
\hat{\mathcal U}_{\text m} \bigl(\ket{g,0} - 2\ket{g,1} + \ket{g,2}\bigr) &\propto \ket{g,\lambda_2},\text{ and}\\
\hat{\mathcal U}_{\text m} \bigl(\ket{g,0} - \ket{g,2}\bigr) &\propto \ket{g,\lambda_3},
\end{aligned}\end{equation}
and states with more than two phonons do not need to be accounted for directly.
Any choice of the motional states $\{\ket{\lambda_i}\}$, with any amount of coherence, is equally efficient under an ideal realisation, as the motion will be completely traced out by the qubit projective measurement.
Similarly, the exact choice of basis of the target-orthogonal space on the left-hand side of eq.~\eqref{eq:measurement-mapping} is arbitrary, as satisfying the equation will result in any orthogonal state being mapped to the qubit state $\ket g$ by linearity.
Beyond this, the efficiency of returning conclusive information from the certifier is dependent on the error characteristics of the particular experiment.
As with entanglement witnesses, the greater the deviation of the given state from the expectation, or the error per coherent operation, the less concrete information the method is likely to be able to return.

A sequence of sideband pulses producing dynamics satisfying eq.~\eqref{eq:measurement-mapping} was found for each target state presented here.
The error in such a map is the total probability that a measurement taken after the map is applied to one of the states in eq.~\eqref{eq:measurement-mapping} would not produce the desired value.
The sequences were found by numerical minimisation of this error over the duration and phase of each pulse in a variety of candidate sequences.
In practice, many sequences exist that result in a probability consistent with zero to a tolerance of \num{e-10}, so the mappings requiring the fewest pulses and the shortest absolute times per shot were chosen to minimise the accumulation of errors from frequency and power drifts.
The first pulse is typically the inverse of the last step of the corresponding creation sequence, reducing the highest-occupied phonon state by one.
Subsequent pulses usually alternate between the carrier and a motion-modifying sideband, such that the phonon count never rises above its initial maximum.

The particular sequence of pulses used to create the target state $\bigl(\ket{g,0} + \ket{g,1} + \ket{g,2}\bigr)/\sqrt3$ is shown in table~\ref{tab:pulses}, along with the subsequent measurement-mapping sequence.
The specifications of sequences for other states are presented in Appendix B, and in machine-readable format in the Supplemental Information~\footnote{See Supplemental Information at \href{https://github.com/ImperialCQD/Certifying-Multilevel-Coherence-in-the-Motional-State-of-a-Trapped-Ion}{https://github.com/ImperialCQD/Certifying-Multilevel-Coherence-in-the-Motional-State-of-a-Trapped-Ion}}.

To produce the interference pattern, a period of phase evolution $\hat{\mathcal U}_{\text f}(\phi)$ under the system Hamiltonian eq.~\eqref{eq:system-hamiltonian} must be implemented.
In any reference frame, this evolution manifests itself as a phase offset to the applied laser field producing the interactions; one interpretation is that the laser phase advancement is paused while it is not interacting with the ion.
In practice one can apply an arbitrary phase offset to the laser field, allowing any free-evolution phase-accumulation to be applied in constant time.
When applying a free-evolution phase of $\phi$, all following red-sideband pulses are phase-shifted by $-\phi$, while all blue-sideband pulses are offset by $\phi$, and the carrier is untouched.
This free-evolution period is imposed between the state creation and measurement-mapping sequences, so only the final five pulses in table~\ref{tab:pulses} are affected, for example.
This is the same operation as pausing the laser evolution, except for time-dependent noise processes which are relevant to determining the duration for which coherence remains, but not to the certification of its initial presence.

\section{Experimental Results}
\label{sec:results}

\begin{figure}
    \includegraphics[width=\linewidth]{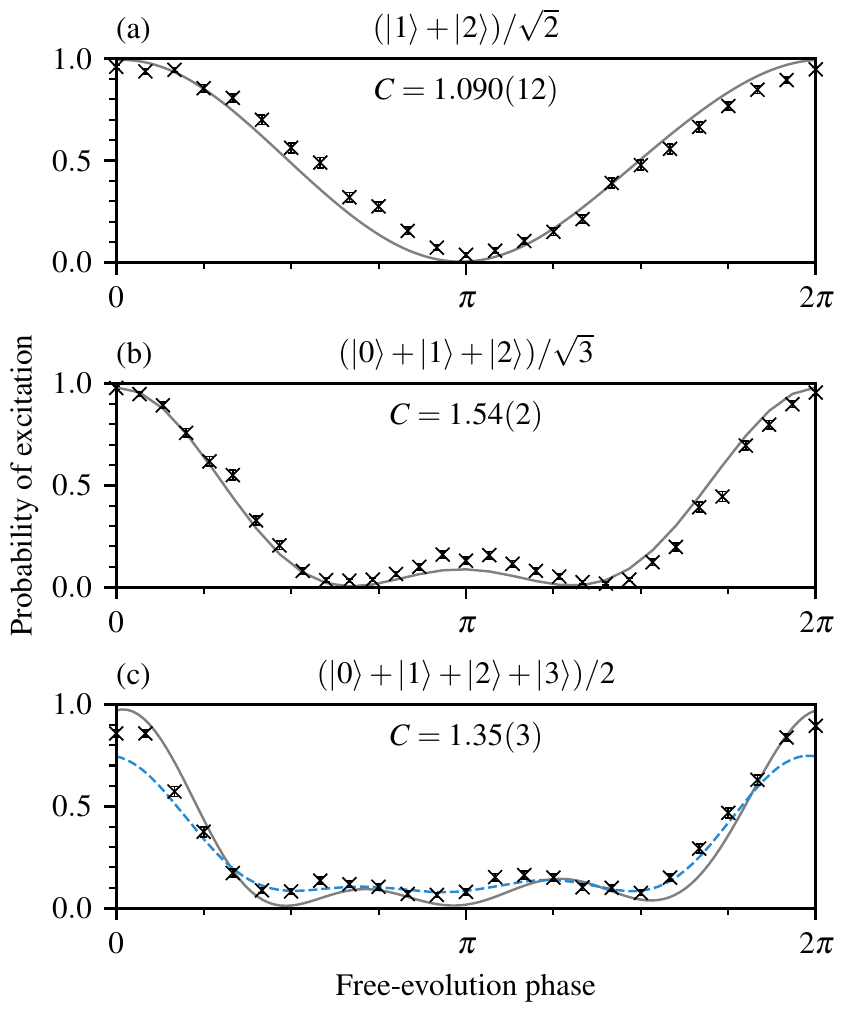}
    \caption{\label{fig:phase}%
    Measured interference patterns (black crosses) for the motional states indicated, using the full measurement-mapping sequences calculated to give a theoretically maximal pattern visibility.
    The models (solid grey line) are numerical simulations using experimental parameter values measured during the trap calibration.
    In all cases, the target also has the qubit in the ground $\ket g$ state.
    The typical Rabi frequency on the carrier was $\SI{90}{\kilo\hertz}$, and 400 shots were taken per plotted data point.
    The depicted probability errors are from the standard estimators of binomial distributions, and are mostly contained within the drawn points.
    The values of $C$ and their uncertainties are determined through numerical integration of the moments of the patterns, with statistical corrections to account for estimator bias.
    (a): the value of $C$ is greater than the threshold needed to unambiguously certify 2-coherence, although the clear oscillation would be a sufficient indicator regardless.
    (b): the value of $C$, \num{1.54(2)}, is sufficient to verify that the created state had least three coherent superposition elements.
    (c): the created state was intended to be 4-coherent, however the value of $C = \num{1.35(3)}$ is only great enough to verify a 3-coherent state.
    The blue dashed line is a pattern from the 3-coherent state that best approximates the data while matching the known state populations from the blue-sideband scan, found numerically. The unconvincing fit suggests that the underlying state may well have been 4-coherent, even though the $C$-test was inconclusive.
}
\end{figure}

The two-element superposition state $(\ket{g,1}+\ket{g,2})/\sqrt2$ was used to initially demonstrate the procedures for preparation and coherence certification of arbitrary motional states.
This state was chosen as it is the lowest-excitation two-element superposition state that requires basis-element populations to interact coherently during the creation, and that does not give a full-visibility interference pattern when the mapping operation is the inverse of the creation.
For an idealised realisation of this state, this na\"ive choice of measurement-mapping sequence can only produce an interference pattern with a theoretical maximum peak-to-peak visibility of $\approx0.88$.
Instead, the more rigorous measurement-mapping sequences described in section~\ref{sec:implementation-certification} were used to give the full-visibility pattern seen in fig.~\ref{fig:phase}(a).
The value of $C$ of \num{1.090(12)} is, as expected, less than the 3-coherence threshold of $5/4=1.25$, which no 2-coherent state can exceed.
After preparing the state using the method of section~\ref{sec:implementation}, the populations of $\ket{g,1}$ and $\ket{g,2}$ were estimated to be \SI{54.7(14)}{\percent} and \SI{38.0(13)}{\percent}, respectively, with \SI{7.2(16)}{\percent} outside these targeted basis elements.
These population numbers and associated 1-$\sigma$ confidence bounds were found by maximum-likelihood estimation using the blue-sideband-probe technique.
The population outside the desired basis elements caused the visibility of the measured interference pattern to be slightly reduced, which, along with other imperfections in implementation, prevented the derived value of $C$ from reaching its maximal value.

Evidence of any oscillation is sufficient to certify 2-coherence, allowing verification even with poor choices of measurement-mapping sequences.
This is not the case for higher-order coherence.
For the state $(\ket{g,0}+\ket{g,1}+\ket{g,2})/\sqrt3$, if the mapping is again the inverse of the state-creation sequence, the certifier has a theoretical maximal value of $C\approx0.92$.
This is significantly below the threshold to certify 3-coherence, in part because the peak-to-peak visibility of the pattern could not exceed $\approx0.68$.
Instead, the optimised measurement mapping specified in table~\ref{tab:pulses} was used, with the resulting interference pattern shown in fig.~\ref{fig:phase}(b).
This gave a measured value of $C = \num{1.54(2)}$, above the threshold of 1.25, unequivocally certifying the state as 3-coherent in the motional basis.
The maximal value, produced with perfect creation and measurement mapping, is $47/27\approx1.74$.
This particular measurement illustrates the robustness of the metric to imperfect measurement-mapping operations.
Immediately after state preparation, the populations in the three targeted basis elements $\ket{g,0}$, $\ket{g,1}$ and $\ket{g,2}$ were \SI{33(2)}{\percent}, \SI{30(2)}{\percent} and \SI{33(2)}{\percent}, respectively, and the remaining \SI{4.7(14)}{\percent} was in other states only intended to be populated during the creation algorithm.
These values were derived from the data in fig.~\ref{fig:bsb}.

The solid grey curves in fig.~\ref{fig:phase} are from a numerical simulation of the state creation and certification process.
These simulations include the effects of off-resonant driving of the carrier when sideband transitions are addressed, which reduce the value of $C$ expected for the $3$-coherent state to \num{1.69} for the experimental parameters used.
This model is not however statistically consistent with the measured data at the 1-$\sigma$ level, most likely due to miscalibration and drift of parameters over the course of the experiment, as well as non-ideal initial preparation of the motional ground state.
The certifier is especially sensitive to detuning errors: for the carrier Rabi frequency of $\SI{90}{\kilo\hertz}$ used in the experiment, a miscalibration of the carrier or trap frequencies by less than $\SI{1}{\kilo\hertz}$ would reduce the maximum value obtainable by the certifier below the 3-coherence threshold.
Alternatively, an initial thermal state defined by $\bar{n} = 0.02$ would reduce $C$ to $1.65$ in the absence of parameter errors.
Despite this demonstrable presence of uncontrolled imperfections, the certifier was still able to  certify unambiguously that 3-coherence had been created.

It is to be expected that higher orders of coherence are more difficult to create and certify.
Figure~\ref{fig:phase}(c) shows the interference pattern resulting from the attempted creation and measurement of an equal, in-phase 4-element superposition of $\ket{g,0}$ through $\ket{g,3}$.
The measured interference-pattern data only achieved $C = \num{1.35(3)}$, below the 4-coherence threshold of $179/96\approx1.86$.
This is an inconclusive result; it is possible that the state was 4-coherent immediately after creation, but imperfections in the mapping meant that this went unverified.
The populations of the four basis elements were \SI{29(2)}{\percent}, \SI{25(2)}{\percent}, \SI{21(2)}{\percent} and \SI{22(2)}{\percent}, respectively, with \SI{3.7(14)}{\percent} outside the expected-state subspace.
The nature of the creation sequences requires some degree of coherence to be present in order to achieve such population numbers, offering evidence in support of this hypothesis.
To further illustrate this key point, the dashed blue line in fig.~\ref{fig:phase}(c) is a simulated interference pattern resulting from the same mapping, implemented perfectly, applied to a 3-coherent motional state consistent with the estimated populations.
The state was chosen to maximise the likelihood of measuring the given data, yet the resulting fit is not very convincing.
This mismatch to the measured pattern shape is also suggestive that the underlying state was, in fact, 4-coherent, but a sufficient build-up of imperfections led to the moment ratio $C$ being less than the certification threshold.
If one is prepared to relax the tolerance for certification from ``beyond reasonable doubt'' to a model-dependent ``balance of probability'', this maximum-likelihood estimation extends the analysis without additional experimental cost.
Exceeding a threshold with the certifier $C$ still, however, meets the former standard of proof, as was the case for the certification of 3-coherence in both this and the previous states.

\section{Discussion}

In moving from a theoretical description of the coherence certifier to a physical realisation, it is important to recognise the potential violations of assumptions, to ensure that the fail-safe nature of the metric is not contravened.
Imperfections in realising the mapping operation $\hat{\mathcal U}_{\text m}$ cannot result in measuring a value of $C$ in excess of the relevant threshold for any input state, provided that such imperfections are independent of the free-evolution phase $\phi$~\cite{Dive2020}.
It is worth considering how such a phase-dependent error might enter during the experiment, in order to be confident that no such effect has impacted the measurements.
To illustrate with an extreme example, an incoherent state $\ket0$ could be incorrectly measured as 2-coherent if the implemented mapping created a projection on to an evolution-phase-dependent state $\cos\phi\ket0 + \sin\phi\ket1$, or the state-distinction measurement had a time-dependent accuracy.

The particular method of effecting the free-evolution period by applying phase shifts to the laser fields eliminated several classes of time-dependent error.
With no physical wait time, the duration of every shot of a given interference pattern was equal, preventing time-dependent drifts in controls from introducing extra features to the patterns.
The free-evolution phase shifts were all added by the arbitrary waveform generator as part of the same procedure as fixing the phase of a sideband pulse in sequence, and consequently any error is independent of the magnitude of the intended shift.
All other time-dependent variations were addressed by randomising the order in which the data points were taken.
Each data point depicted in fig.~\ref{fig:phase} was derived from 400 individual shots.
To create a pattern, four rasters through each set of free-evolution phases were taken, each providing 100 shots to every point, with the order of the phases randomised for each raster.
Approximately, this converts any possible periodic drifts into incoherent white-noise processes spread evenly across the measurements.

While these particular steps severely limit the possibility of drift artificially inflating the measured value of the certifier, they do not affect systematic miscalibrations of the various transition frequencies.
Severe but stable deviations from the two sideband frequencies could, in principle, lead to a phase dependence of the measurement mapping.
These could arise either from imperfect compensation of the AC Stark effect, or from systematic mis-sets of the laser frequencies.
The most noticeable effects of detuning from a sideband by an amount $\delta$ are a modification of the sideband Rabi frequency $\Omega'$ and a reduction in Rabi-flop visibility, both by a factor of $\sqrt{1 + \delta^2/\Omega'^2}$.
This timing error results in the free-evolution phase-advancement operations applying two different erroneous phases to the sidebands, producing an incorrect mapping operation that depends on the amount of phase to be applied.

This can be detected principally by observing sideband probes similar to fig.~\ref{fig:bsb} taken immediately after ground-state cooling with less-than-unity visibility. 
Frequency fluctuation measurements in the trap suggested deviations in the sideband calibration frequency were less than $\delta/\Omega' = \num{0.15}$ over the course of one experimental run.
However, the time-dependent components of this drift were converted to incoherent processes by the aforementioned point randomisation, which in simulations completely dominated any remnant effects that might have overestimated the value of $C$.
Similarly, any possible contributions from the small-amplitude high-frequency oscillations stemming from the nearby carrier transition are rendered incoherent by the drift of the relevant parameters, and consequently cannot cause $C$ to increase beyond a threshold.

A final consideration is the effect of incomplete statistics on the estimator of $C$.
The individual shots of data making up the interference pattern are samples from independent Bernoulli distributions, and the estimator $\hat C$ and its uncertainty must be derived from these.
A na\"ive estimator constructed by a direct discretisation of the integrals in the definition of the moments leads to
\begin{equation}\label{eq:c-estimator}
\hat C_{\text{na\"ive}} = \frac{\sum_i w_i {(\bar p_i)}^3}{{\bigr(\sum_i w_i \bar p_i\bigr)}^2},
\end{equation}
for numerical integration weights $\{w_i\}$ and excited-state probability estimators $\{\bar p_i\}$.
The integration weights chosen are predetermined, and depend only on the quadrature method to be used and the locations of the sampled free-evolution phases.
However, the nonlinear dependence of this estimator on the $\bar p_i$ results in it having a systematic bias upwards; for perfect observations of a state that achieves a threshold value of $C$, the estimated value $\hat C_{\text{na\"ive}}$ would be more likely than not to produce a value that was too high, and this increase would not be accounted for by a simple lowest-order propagation of uncertainty.
Instead, the estimator of $C$ and its standard deviation used in this work are derived from eq.~\eqref{eq:c-estimator} with additional bias-correction terms calculated from the method of moments~\cite{Casella2002} using estimators of the binomial distribution's central moments up to and including the skewness~\cite{Chan2020}.
At 400 shots-per-point, any remaining bias was found by Monte-Carlo simulation to be at an order of magnitude smaller than the quoted uncertainties.
Full details of the statistical estimators used are presented in Appendix C.
% Note: doesn't use \ref because revtex doesn't support ref'ing appendices.  Be sure to change if we change appendices.

\section{Conclusion}
\label{sec:conclusion}

Multilevel coherence can be unambiguously certified by interference-pattern experiments which vary a single parameter, even in cases where the physical system of interest is not accessible to direct measurement.
These methods only require an available measurement on any coupled system, and are robust against imperfections in the implementation of coherent manipulations mapping the target system to the measurement basis.
With the exception of poor measurement statistics, they cannot produce a false positive.
In cases where the certification test is inconclusive at a particular level, additional maximum-likelihood estimation can be used to indicate whether it is probable that this degree of coherence was achieved, still based only on the interference pattern.

The motion of a single trapped ion can be controlled using only first-order sideband pulses to create nonclassical states that exhibit multilevel coherence.
This can be unambiguously verified by measurements made only of the associated qubit system, similarly using only first-order coupling transitions.
The simplicity of these required operations demonstrate the applicability of the coherence-certification method to any physical system where some elements are slow, unreliable or unfeasible to measure.
This includes any set of entangled qudits where some of the systems may not be accessed by interrogation fields, such as in cavity optomechanics or superconducting qubits coupled to a resonator.

Verification of the entanglement and coherence properties of arbitrary quantum states is experimentally taxing, often needing full reconstruction of the density operator.
This requires a level of control beyond current noisy intermediate-scale quantum devices.
The problem is exacerbated when some subsystems are inaccessible to measurement, subjecting tomographic procedures to large errors, and they are rendered completely unreliable when coherent manipulations of the system cannot be trusted.
The resilient interference-pattern metrics presented here do not suffer such limitations.
The certification is read from a simple, single statistic of a data set found by varying a single parameter, and is consequently far more forgiving of statistical uncertainty than those based on decompositions of a matrix where each element has a large confidence interval.

Despite decoherence being a major roadblock to large-scale quantum computing, the intricacies of high-order coherence remain little understood.
The uses of coherence as a resource are also still being actively investigated for their roles in thermodynamic and solid-state transport processes.
These interference-pattern metrics provide an experimentally realistic method to effect these analyses, paving the way to a greater understanding of the fundamental nature of quantum mechanics.

\begin{acknowledgments}
We are immensely grateful to Brian Willey for his role in the fabrication of the trap and other laboratory equipment.
% general ion-trap grant
Financial support by EPSRC through all of \textit{Optimal control for robust ion trap quantum logic} Grant No.\ EP/P024890/1,
% hub
\textit{Hub in Quantum Computing and Simulation} Grant No. EP/T001062/1,
% CDT
the \textit{Centre for Doctoral Training in Controlled Quantum Dynamics} Grant No.\ EP/L016524/1,
% Ollie and Jake
and the \textit{Training and Skills Hub in Quantum Systems Engineering} Grant No.\ EP/P510257/1 is gratefully acknowledged.
\end{acknowledgments}

\clearpage

\appendix
\makeatletter\DeclarePairedDelimiter\@expectfences{[}{]}\newcommand*\expect{\operatorname E\@expectfences}\makeatother

\section{Numerical Determination of Threshold Values}

The threshold values when using generalised measurements were estimated by numerically maximising the value of the certifier over the space of $k$-coherent states, and general measurement operators.
This can be achieved by a general-purpose unconstrained optimisation routine such as the common Broyden--Fletcher--Goldfarb--Shanno algorithm by finding a parametrisation function that takes an input vector in $\mathbb R^\ell$ to the search space, for some arbitrary number of parameters $\ell$.
Any POVM value $\hat{\mathcal A}$ may be considered a sum of simple projective measurements, such that $\hat{\mathcal A} = \sum_j a_j \ket{\psi_j}\!\bra{\psi_j}$ for constants $0\le a_j\le1$, and a set of orthonormal basis states $\{\ket{\psi_j}\}$.
Any $k$-coherent density matrix can be written as $\hat\rho = \sum_j p_j \hat\rho_j$, where here $\sum_jp_j = 1$ and the $\{\hat{\rho_j}\}$ are density matrices each in a $k$-dimensional subspace of the full Hilbert space.
A suitable parametrisation can be derived from these two forms.
For convenience, we optimise separately over different numbers of nonzero $a_j$ and $p_j$.
The case of having only one each of these is the same as the problem considered in ref.~\cite{Dive2020}.

\subsection{Methods of Parametrisation}

The parametrisation of a POVM value with $m$ projective components can be done iteratively.
First, parametrise a pure state in the complete Hilbert space of the measurement.
Second, calculate the subspace orthogonal to this chosen state.
Next, parametrise another pure state, this time out of the subspace, and calculate the new subspace orthogonal to both selected states.
Repeat this process until the desired number of components has been found.
This gives $m$ orthonormal pure states; one need now only parametrise $m$ values between $0$ and $1$ to be the $a_j$, and then calculate $\hat{\mathcal A}$ from the pieces.

To generate a valid pure state in an $n$-dimensional Hilbert space, one can generate $n-1$ amplitudes $c_q$ and phases $\theta_q$, then return the normalised dot product of the vector $(1,\,c_1e^{i\theta_1},\,c_2e^{i\theta_2},\,\dotsc)$ with an arbitrary basis of the Hilbert space.
The orthogonal subspace can be found by applying the Gram--Schmidt process to an initial spanning set of states, which will also produce a suitable orthogonal basis to use in the next step.
The constraint on the individual $a_j$ values in the POVM value can be handled by any standard method, such as logit transformations.
With an optimiser, unlike with random sampling, is it generally not necessary to ensure that the output is distributed according to the Haar measure, and so this biased parametrisation is only unsuitable if the optimisation landscape becomes too flat for the routine to converge.

For the $k$-coherent density matrix in a larger Hilbert space, one can take the subspaces in the sum as being those spanned by all combinations of drawing $k$ vectors from a full-space basis.
Within each subspace, there is no constraint on the allowable density matrices.
All density matrices have a Cholesky decomposition $\hat\rho = LL^\dagger$ for a lower-triangular matrix $L$.
An $n$-dimensional density matrix can therefore be found by taking $n(n+1)/2$ parameters to be the magnitudes of the entries of some related $L'$, and using a further $n(n-1)/2$ parameters to be the phases of the off-diagonal elements.
The desired density matrix is then $\hat\rho = L'L'^\dagger / \Tr(L'L'^\dagger)$.
The necessary probabilities $p_j$ assigned to the individual density matrices can then be parametrised similarly to the $a_j$ values, except one fewer parameter is needed; their sum must be scaled to be equal to one, so only the relative amplitudes are required.

\subsection{Results}

Optimisations were run in Hilbert spaces of varying sizes, with varying ranks of the measurement operator $\hat{\mathcal A}$ and subspaces of measured state $\hat\rho$ (corresponding to the number of nonzero $a_j$ and $p_j$ respectively).
In all cases, the optimiser would reliably attempt to reduce the density matrix to a single pure state, and reduce the measurement operator to be a simple projective measurement onto the same state.
The states found were entirely consistent with the results of ref.~\cite{Dive2020}.

Further, forcing the measurement operator to have more than one $j$ satisfy $a_j = 1$ meant that the optimiser could only achieve one threshold lower than the best case, for every additional rank added above the first.
For example, a measurement operator of the form $\ket{\psi_1}\!\bra{\psi_1} + \ket{\psi_2}\!\bra{\psi_2}$ in a 4-dimensional Hilbert could never measure more than 3-coherence.
This seems intuitive; adding extra orthogonal components to the measurement operator means that less of the Hilbert space is distinct from the measured state, and consequently one expects less variation in the interference pattern.
Since the certifier is related to the ratio of deviations in the pattern to its average, it is not unexpected that these higher-rank measurements would only give worse results.
This is not expected to be a problem in the present experiment, as the methods described in section~\ref{sec:implementation-certification} specifically map only the target state into the measurement subspace, and all other relevant states into the measurement-orthogonal space.

\clearpage

\begin{widetext}

\section{Pulse Sequence for all States}

This appendix tabulates the pulse sequences used for the state creation and measurement mapping for each of the states presented in the main text, in tables~\ref{tab:pulses-12} to \ref{tab:pulses-0123}.
The pulse length is scaled relative to the oscillation frequency of the coupled pair that includes the motional $\ket0$ state, so that a value of $1$ is the time taken to exchange $\ket{g,0}\leftrightarrow\ket{e,0}$ on the carrier and $\ket{g,1}\leftrightarrow\ket{e,0}$ on the red sideband.
The given phase is applied as an offset, so that the set laser phase at the beginning of a pulse is offset relative to where it would have been had it oscillated freely at its transition frequency since the beginning of the experiment.
The interference pattern is constructed by varying the phase offsets of all blue- and red-sideband pulses during the measurement mapping sequences only.
The phase offset should be added to blue-sideband pulses, and subtracted from red-sideband pulses.
\begin{table*}%
    \newcolumntype{x}{D..{1.2}}%
    \begin{ruledtabular}\begin{tabular}{l@{\hskip 1.5em}xxxxc@{\hskip 1.5em}xxxxx}
    & \multicolumn{4}c{State creation} && \multicolumn{5}c{Measurement mapping}\\
        \rule{0pt}{2.5ex}Transition & \multicolumn1c{carrier} & \multicolumn1c{red} & \multicolumn1c{carrier} & \multicolumn1c{red} && \multicolumn1c{red} & \multicolumn1c{carrier} & \multicolumn1c{red} & \multicolumn1c{carrier} & \multicolumn1c{red}\\
        Pulse length          &  0.60 &  0.80 &  0.74 &  0.71 &&  0.71 &  0.44 &  1.41 &  0.54 &  1.41\\
        Phase offset ${}/\pi$ &  0    & -0.50 &  0    & -0.50 &&  0    & -0.66 & -0.83 & -0.87 & -0.41
    \end{tabular}\end{ruledtabular}%
    \caption{\label{tab:pulses-12}%
        Pulse sequences for creation and measurement mapping of target state $\bigl(\ket{g,1} + \ket{g,2}\bigr)/\sqrt{2}$.
    }%
\end{table*}
\begin{table*}%
    \newcolumntype{x}{D..{1.2}}%
    \begin{ruledtabular}\begin{tabular}{l@{\hskip 1.5em}xxxxc@{\hskip 1.5em}xxxxx}
    & \multicolumn{4}c{State creation} && \multicolumn{5}c{Measurement mapping}\\
        \rule{0pt}{2.5ex}Transition & \multicolumn1c{carrier} & \multicolumn1c{red} & \multicolumn1c{carrier} & \multicolumn1c{red} && \multicolumn1c{red} & \multicolumn1c{carrier} & \multicolumn1c{red} & \multicolumn1c{carrier} & \multicolumn1c{red}\\
        Pulse length          &  0.50 &  0.70 &  0.73 &  0.71 &&  0.71 &  0.50 &  1.42 &  1.59 &  0.72\\
        Phase offset ${}/\pi$ &  0    & -0.50 &  1.00 &  0.50 &&  0    &  0.71 & -0.29 &  0.10 & -0.51
    \end{tabular}\end{ruledtabular}%
    \caption{\label{tab:pulses-012}%
        Pulse sequences for creation and measurement mapping of target state $\bigl(\ket{g,0} + \ket{g,1} + \ket{g,2}\bigr)/\sqrt{3}$.
    }%
\end{table*}
\begin{table*}%
    \newcolumntype{x}{D..{1.2}}%
    \begin{ruledtabular}\begin{tabular}{l@{\hskip 1.5em}xxxxxxc@{\hskip 1.5em}xxxxxxxxx}
    & \multicolumn{6}c{State creation} && \multicolumn{9}c{Measurement mapping}\\
        \rule{0pt}{2.5ex}Transition & \multicolumn1c{carrier} & \multicolumn1c{red} & \multicolumn1c{carrier} & \multicolumn1c{red} & \multicolumn1c{carrier} & \multicolumn1c{red} && \multicolumn1c{red} & \multicolumn1c{carrier} & \multicolumn1c{blue} & \multicolumn1c{carrier} & \multicolumn1c{red} & \multicolumn1c{carrier} & \multicolumn1c{red} & \multicolumn1c{carrier} & \multicolumn1c{red}\\
        Pulse length          &  0.51 &  0.55 &  0.96 &  0.57 &  0.84 &  0.58 &&  2.89 &  1.47 &  1.15 &  3.02 &  2.31 &  4.69 &  2.31 &  0.72 &  0.58\\
        Phase offset ${}/\pi$ &  0    & -0.50 & -1.00 &  0.50 &  0    & -0.50 &&  0    & -0.16 & -0.41 & -0.53 &  0.45 &  0.79 & -0.32 & -0.13 &  0.76
    \end{tabular}\end{ruledtabular}%
    \caption{\label{tab:pulses-0123}%
        Pulse sequences for creation and measurement mapping of target state $\bigl(\ket{g,0} + \ket{g,1} + \ket{g,2} + \ket{g,3}\bigr)/2$.
    }%
\end{table*}

\clearpage

\section{Statistics of the Moment Ratio}
\label{sec:statistics-ratio}

\subsection{Estimating the Moment Ratio}

Each point of an interference pattern is an independent binomial distribution with some exact underlying probability $\mu_j$, which is estimated by taking a number of shots $n$ and counting the number of ``successes''.
The estimates $p_j$ are unbiased estimates of the means of the distributions.
The quantity of interest is
\begin{equation}
\bar c = \frac{m_3}{m_1^2} = \frac{\frac1{2\pi}\int_0^{2\pi} p(\phi)^3\,d\phi}{{\Bigl(\frac1{2\pi}\int_0^{2\pi}p(\phi)\,d\phi\Bigr)}^2},
\end{equation}
using a finite number of points ($J$, the number of $p_j$) and a finite number of shots.
The interference patterns vary sufficiently smoothly that the trapezium rule is appropriate, so
\begin{equation}
\frac1{2\pi}\int_0^{2\pi} f(x)\,dx \approx \sum_{j=0}^{J-1} w_j f\Bigl(\frac{2\pi j}{J-1}\Bigr),
\quad\text{where\ }
w_j = \begin{cases}\frac1{2(J-1)}&\text{$j=0$ or $j=J-1$}\\\frac1{J-1}&\text{all other $j$.}\end{cases}
\end{equation}

Let $P_j \sim \mathrm B(n, \mu_j)/n$ be the distribution from which each of the $p_j$ estimates are drawn.
Similarly, let
\begin{equation}
    C \sim \frac{\sum_j w_j P_j^3}{{\Bigl(\sum_j w_j P_j\Bigr)}^2}
\end{equation}
be the distribution of the estimates $c$ of the true value $\bar C$.

The expectation of a measurement of $C$ is $\expect{C}$ where the expectation runs over all the independent probability distributions $P_j$.
For each point's distribution the expectation is $\expect{P_j} = \expect[\big]{\mu_j + (P_j - \mu_j)}$.
The unusual form is to permit a Taylor expansion of the expectation of $C$ around the binomial means in terms of $\expect[\big]{{(P_j - \mu_j)}^n}$, the central moments of the distributions.
For the binomial distribution, unbiased estimators (symbols with hats) of these are:
\begin{equation}\label{eq:moments}\begin{alignedat}3
\expect[\big]{{(P_j - \mu_j)}^1} &\to p_j - \hat\mu_j &&= 0 &\quad&\text{(mean)}\\
\expect[\big]{{(P_j - \mu_j)}^2} &\to \hat \sigma_j^2 &&= \frac{p_j (1 - p_j)}{n-1} &&\text{(variance)}\\
\expect[\big]{{(P_j - \mu_j)}^3} &\to \hat \kappa_j &&= \frac{p_j (1 - p_j) (1 - 2p_j)}{(n-1)(n-2)} &&\text{(skewness)}.
\end{alignedat}\end{equation}

The expansion, up to terms of third order, is
\begin{equation}\label{eq:expectation-r3}\begin{aligned}
    \expect{C} &= \expect[\bigg]{\Bigl(\sum_j w_j P_j^3\Bigr){\Bigl(\sum_j w_j P_j\Bigr)}^{-2}}\\
    &\approx \frac{\tilde m_3}{\tilde m_1^2}
        + \underbrace{\frac1{\tilde m_1^2}\sum_j\Biggl(
            3w_j\Bigl(\mu_j - \frac2{\tilde m_1}\mu_j^2 + \frac{\tilde m_3}{\tilde m_1^2}\Bigr)\expect[\big]{{(P_j - \mu_j)}^2}
            + w_j\Bigl(1 - \frac6{\tilde m_1^2}w_j\mu_j + \frac9{\tilde m_1^2}w_j^2\mu_j^2 - 4\frac{\tilde m_3}{\tilde m_1^3}w_j^2\Bigr)\expect[\big]{{(P_j - \mu_j)}^3}
    \Biggr)}_{\text{bias term}},
\end{aligned}\end{equation}
where $\tilde m_1 = \sum_j w_j \mu_j$ and $\tilde m_3 = \sum_j w_j \mu_j^3$.
The desired estimator is $\hat c = \tilde m_3/\tilde m_1^2$, so to make a fair estimator the bias term in eq.~\eqref{eq:expectation-r3} must be subtracted from the ``direct'' measurement $c$.

\subsection{Estimating Variance}

The variance is well approximated in this case by the low-order expansion
\begin{equation}
\sigma_{c} \approx \sqrt{\sum_j {\biggl\lvert \frac{\partial \hat c}{\partial p_j} \biggr\rvert}^2 \sigma_{p_j}^2}\,,
\end{equation}
as direct measurements ($c$) of $C$ are approximately normally distributed, and there is no covariance between the $p_j$.
The estimator $\hat c = c - \sum_j z_{2j} - \sum_j z_{3j}$, where $z_{nj}$ is the term in eq.~\eqref{eq:expectation-r3} including $\expect[\big]{(P_j - \mu_j)^n}$.
The derivatives are then
\begin{equation}\begin{aligned}
\frac{\partial z_{2k}}{\partial p_j} ={}
&\frac{p_k(1-p_k)}{n-1} \Biggl(
   \frac{8w_jw_k^2p_k^2}{\tilde m_1^4}
    + \frac{9w_jw_k^2p_j^2}{\tilde m_1^4}
    - \frac{6w_jw_kp_k}{\tilde m_1^3}
    - \frac{12w_jw_k^2\tilde m_3}{\tilde m_1^5}
\Biggr)\\
&{}+\delta_{jk} \Biggl[
    \frac{p_k(1-p_k)}{n-1}\biggl(
        \frac{3w_k}{\tilde m_1^2} - \frac{12w_k^2p_k}{\tilde m_1^3}
    \biggr)
    + \frac{1-2p_k}{n-1}\cdot\frac{3w_k}{\tilde m_1^2}\biggl(
        p_k - \frac{2w_kp_k^2}{\tilde m_1} + \frac{\tilde m_3w_k}{\tilde m_1^2}
    \biggr)
\Biggr]
\end{aligned}\end{equation}
for the second-order correction terms, and
\begin{equation}\begin{alignedat}2
\frac{\partial z_{3k}}{\partial p_j} = {}
&\frac{p_k(1-p_k)(1-2p_k)w_j}{(n-1)(n-2)} \Biggl(
    - \frac{2w_k}{\tilde m_1^3}
    + \frac{20\tilde m_3w_k^3}{\tilde m_1^6}
    + \frac{24w_k^2p_k}{\tilde m_1^4}
    - \frac{36w_k^3p_k^2}{\tilde m_1^5}
    - \frac{12w_k^3p_j^2}{\tilde m_1^3}
\Biggr)\span\span\\
&{}+\delta_{jk} \frac{w_k}{\tilde m_1^2} \Biggl[
    &&\biggl(
        \frac{18w_k^2p_k}{\tilde m_1^2}
        - \frac{6w_k}{\tilde m_1^2}
    \biggr)\frac{p_k(1-p_k)(1-2p_k)}{(n-1)(n-2)}
    + \biggl(
        1
        - \frac{6w_kp_k}{\tilde m_1^2}
        + \frac{9w_k^2p_k^2}{\tilde m_1^2}
        - \frac{4\tilde m_3w_k^2}{\tilde m_1^3}
    \biggr)\frac{6p_k^2 - 6p_k + 1}{(n-1)(n-2)}
\Biggr]
\end{alignedat}\end{equation}
for the third-order corrections.

\subsection{Validity of Estimators}

\begin{figure}
    \includegraphics{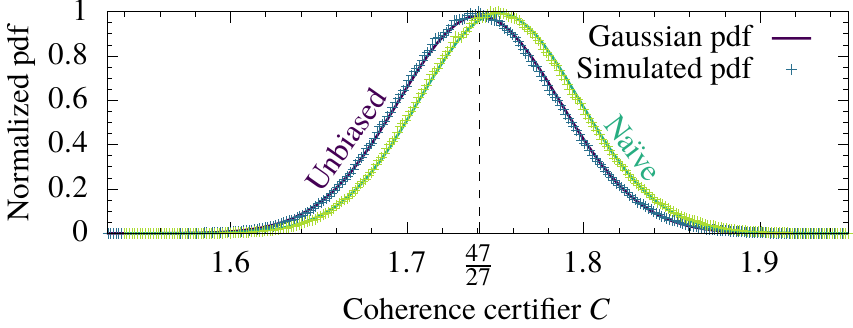}
    \caption{\label{fig:pdf}%
    Normalised probability density functions for the unbiased and biased estimators of the true value $\bar c$ using a 31-point trapezium rule with 100 shots-per-point.
    The plots show direct interpolations of the pdfs (crosses) overlaid on Gaussian approximations (solid lines).
    The Gaussian approximation is excellent, though a very minor skewness is detectable in the direct interpolations.
    The probability distributions were approximated by binning the measured values from one million simulations of measurement of a pattern whose analytic moment ratio $C = 47/27$.
    Each data point in each simulation was found by randomly sampling a binomial distribution with 100 shots at the expected probability.
    The same data sets were used to evaluate both estimators.
    The na\"ive estimator even with 100 shots per point systematically overestimates the true value, whereas the unbiased one is correct.
}
\end{figure}

Figure~\ref{fig:pdf} shows Monte-Carlo approximations to the probability density functions of the two unbiased and na\"ive estimators (crosses) and Gaussian approximations of these distributions, using a 31-point trapezium rule and 100 shots-per-point for an idealised superposition state $\bigl(\ket0+\ket1+\ket2\bigr)/\sqrt3$
There is a small skewness visible in the Monte-Carlo pdfs, which shows the true distribution of $c$ is not exactly Gaussian, though this is a very minor effect.
The plot shows the bias in the na\"ive estimator clearly; an unbiased estimator should have a mean at $47/27$ as marked by the dashed vertical line.

\end{widetext}


\begin{thebibliography}{33}%
\makeatletter
\providecommand \@ifxundefined [1]{%
 \@ifx{#1\undefined}
}%
\providecommand \@ifnum [1]{%
 \ifnum #1\expandafter \@firstoftwo
 \else \expandafter \@secondoftwo
 \fi
}%
\providecommand \@ifx [1]{%
 \ifx #1\expandafter \@firstoftwo
 \else \expandafter \@secondoftwo
 \fi
}%
\providecommand \natexlab [1]{#1}%
\providecommand \enquote  [1]{``#1''}%
\providecommand \bibnamefont  [1]{#1}%
\providecommand \bibfnamefont [1]{#1}%
\providecommand \citenamefont [1]{#1}%
\providecommand \href@noop [0]{\@secondoftwo}%
\providecommand \href [0]{\begingroup \@sanitize@url \@href}%
\providecommand \@href[1]{\@@startlink{#1}\@@href}%
\providecommand \@@href[1]{\endgroup#1\@@endlink}%
\providecommand \@sanitize@url [0]{\catcode `\\12\catcode `\$12\catcode
  `\&12\catcode `\#12\catcode `\^12\catcode `\_12\catcode `\%12\relax}%
\providecommand \@@startlink[1]{}%
\providecommand \@@endlink[0]{}%
\providecommand \url  [0]{\begingroup\@sanitize@url \@url }%
\providecommand \@url [1]{\endgroup\@href {#1}{\urlprefix }}%
\providecommand \urlprefix  [0]{URL }%
\providecommand \Eprint [0]{\href }%
\providecommand \doibase [0]{https://doi.org/}%
\providecommand \selectlanguage [0]{\@gobble}%
\providecommand \bibinfo  [0]{\@secondoftwo}%
\providecommand \bibfield  [0]{\@secondoftwo}%
\providecommand \translation [1]{[#1]}%
\providecommand \BibitemOpen [0]{}%
\providecommand \bibitemStop [0]{}%
\providecommand \bibitemNoStop [0]{.\EOS\space}%
\providecommand \EOS [0]{\spacefactor3000\relax}%
\providecommand \BibitemShut  [1]{\csname bibitem#1\endcsname}%
\let\auto@bib@innerbib\@empty
%</preamble>
\bibitem [{\citenamefont {Oi}\ and\ \citenamefont {{\AA}berg}(2006)}]{Oi2006}%
  \BibitemOpen
  \bibfield  {author} {\bibinfo {author} {\bibfnamefont {D.~K.~L.}\
  \bibnamefont {Oi}}\ and\ \bibinfo {author} {\bibfnamefont {J.}~\bibnamefont
  {{\AA}berg}},\ }\href {https://doi.org/10.1103/PhysRevLett.97.220404}
  {\bibfield  {journal} {\bibinfo  {journal} {Physical Review Letters}\
  }\textbf {\bibinfo {volume} {97}},\ \bibinfo {pages} {220404} (\bibinfo
  {year} {2006})},\ \Eprint {https://arxiv.org/abs/0603157} {arXiv:0603157
  [quant-ph]} \BibitemShut {NoStop}%
\bibitem [{\citenamefont {Girolami}(2014)}]{Girolami2014}%
  \BibitemOpen
  \bibfield  {author} {\bibinfo {author} {\bibfnamefont {D.}~\bibnamefont
  {Girolami}},\ }\href {https://doi.org/10.1103/PhysRevLett.113.170401}
  {\bibfield  {journal} {\bibinfo  {journal} {Physical Review Letters}\
  }\textbf {\bibinfo {volume} {113}},\ \bibinfo {pages} {170401} (\bibinfo
  {year} {2014})},\ \Eprint {https://arxiv.org/abs/1403.2446} {arXiv:1403.2446}
  \BibitemShut {NoStop}%
\bibitem [{\citenamefont {Streltsov}\ \emph {et~al.}(2017)\citenamefont
  {Streltsov}, \citenamefont {Adesso},\ and\ \citenamefont
  {Plenio}}]{Streltsov2017}%
  \BibitemOpen
  \bibfield  {author} {\bibinfo {author} {\bibfnamefont {A.}~\bibnamefont
  {Streltsov}}, \bibinfo {author} {\bibfnamefont {G.}~\bibnamefont {Adesso}},\
  and\ \bibinfo {author} {\bibfnamefont {M.~B.}\ \bibnamefont {Plenio}},\
  }\href {https://doi.org/10.1103/RevModPhys.89.041003} {\bibfield  {journal}
  {\bibinfo  {journal} {Reviews of Modern Physics}\ }\textbf {\bibinfo {volume}
  {89}},\ \bibinfo {pages} {041003} (\bibinfo {year} {2017})}\BibitemShut
  {NoStop}%
\bibitem [{\citenamefont {Levi}\ and\ \citenamefont
  {Mintert}(2014)}]{Levi2014}%
  \BibitemOpen
  \bibfield  {author} {\bibinfo {author} {\bibfnamefont {F.}~\bibnamefont
  {Levi}}\ and\ \bibinfo {author} {\bibfnamefont {F.}~\bibnamefont {Mintert}},\
  }\href {https://doi.org/10.1088/1367-2630/16/3/033007} {\bibfield  {journal}
  {\bibinfo  {journal} {New Journal of Physics}\ }\textbf {\bibinfo {volume}
  {16}},\ \bibinfo {pages} {033007} (\bibinfo {year} {2014})}\BibitemShut
  {NoStop}%
\bibitem [{\citenamefont {Baumgratz}\ \emph {et~al.}(2014)\citenamefont
  {Baumgratz}, \citenamefont {Cramer},\ and\ \citenamefont
  {Plenio}}]{Baumgratz2014}%
  \BibitemOpen
  \bibfield  {author} {\bibinfo {author} {\bibfnamefont {T.}~\bibnamefont
  {Baumgratz}}, \bibinfo {author} {\bibfnamefont {M.}~\bibnamefont {Cramer}},\
  and\ \bibinfo {author} {\bibfnamefont {M.~B.}\ \bibnamefont {Plenio}},\
  }\href {https://doi.org/10.1103/PhysRevLett.113.140401} {\bibfield  {journal}
  {\bibinfo  {journal} {Physical Review Letters}\ }\textbf {\bibinfo {volume}
  {113}},\ \bibinfo {pages} {140401} (\bibinfo {year} {2014})},\ \Eprint
  {https://arxiv.org/abs/1311.0275} {arXiv:1311.0275} \BibitemShut {NoStop}%
\bibitem [{\citenamefont {Ringbauer}\ \emph {et~al.}(2018)\citenamefont
  {Ringbauer}, \citenamefont {Bromley}, \citenamefont {Cianciaruso},
  \citenamefont {Lami}, \citenamefont {Lau}, \citenamefont {Adesso},
  \citenamefont {White}, \citenamefont {Fedrizzi},\ and\ \citenamefont
  {Piani}}]{Ringbauer2018}%
  \BibitemOpen
  \bibfield  {author} {\bibinfo {author} {\bibfnamefont {M.}~\bibnamefont
  {Ringbauer}}, \bibinfo {author} {\bibfnamefont {T.~R.}\ \bibnamefont
  {Bromley}}, \bibinfo {author} {\bibfnamefont {M.}~\bibnamefont
  {Cianciaruso}}, \bibinfo {author} {\bibfnamefont {L.}~\bibnamefont {Lami}},
  \bibinfo {author} {\bibfnamefont {W.~Y.~S.}\ \bibnamefont {Lau}}, \bibinfo
  {author} {\bibfnamefont {G.}~\bibnamefont {Adesso}}, \bibinfo {author}
  {\bibfnamefont {A.~G.}\ \bibnamefont {White}}, \bibinfo {author}
  {\bibfnamefont {A.}~\bibnamefont {Fedrizzi}},\ and\ \bibinfo {author}
  {\bibfnamefont {M.}~\bibnamefont {Piani}},\ }\href
  {https://doi.org/10.1103/PhysRevX.8.041007} {\bibfield  {journal} {\bibinfo
  {journal} {Physical Review X}\ }\textbf {\bibinfo {volume} {8}},\ \bibinfo
  {pages} {041007} (\bibinfo {year} {2018})},\ \Eprint
  {https://arxiv.org/abs/1707.05282} {arXiv:1707.05282} \BibitemShut {NoStop}%
\bibitem [{\citenamefont {Hillery}(2016)}]{Hillery2016}%
  \BibitemOpen
  \bibfield  {author} {\bibinfo {author} {\bibfnamefont {M.}~\bibnamefont
  {Hillery}},\ }\href {https://doi.org/10.1103/PhysRevA.93.012111} {\bibfield
  {journal} {\bibinfo  {journal} {Physical Review A}\ }\textbf {\bibinfo
  {volume} {93}},\ \bibinfo {pages} {012111} (\bibinfo {year}
  {2016})}\BibitemShut {NoStop}%
\bibitem [{\citenamefont {Shi}\ \emph {et~al.}(2017)\citenamefont {Shi},
  \citenamefont {Liu}, \citenamefont {Wang}, \citenamefont {Yang},
  \citenamefont {Yang},\ and\ \citenamefont {Fan}}]{Shi2017}%
  \BibitemOpen
  \bibfield  {author} {\bibinfo {author} {\bibfnamefont {H.-L.}\ \bibnamefont
  {Shi}}, \bibinfo {author} {\bibfnamefont {S.-Y.}\ \bibnamefont {Liu}},
  \bibinfo {author} {\bibfnamefont {X.-H.}\ \bibnamefont {Wang}}, \bibinfo
  {author} {\bibfnamefont {W.-L.}\ \bibnamefont {Yang}}, \bibinfo {author}
  {\bibfnamefont {Z.-Y.}\ \bibnamefont {Yang}},\ and\ \bibinfo {author}
  {\bibfnamefont {H.}~\bibnamefont {Fan}},\ }\href
  {https://doi.org/10.1103/PhysRevA.95.032307} {\bibfield  {journal} {\bibinfo
  {journal} {Physical Review A}\ }\textbf {\bibinfo {volume} {95}},\ \bibinfo
  {pages} {032307} (\bibinfo {year} {2017})},\ \Eprint
  {https://arxiv.org/abs/1610.08656} {arXiv:1610.08656} \BibitemShut {NoStop}%
\bibitem [{\citenamefont {Castellini}\ \emph {et~al.}(2019)\citenamefont
  {Castellini}, \citenamefont {{Lo Franco}}, \citenamefont {Lami},
  \citenamefont {Winter}, \citenamefont {Adesso},\ and\ \citenamefont
  {Compagno}}]{Castellini2019}%
  \BibitemOpen
  \bibfield  {author} {\bibinfo {author} {\bibfnamefont {A.}~\bibnamefont
  {Castellini}}, \bibinfo {author} {\bibfnamefont {R.}~\bibnamefont {{Lo
  Franco}}}, \bibinfo {author} {\bibfnamefont {L.}~\bibnamefont {Lami}},
  \bibinfo {author} {\bibfnamefont {A.}~\bibnamefont {Winter}}, \bibinfo
  {author} {\bibfnamefont {G.}~\bibnamefont {Adesso}},\ and\ \bibinfo {author}
  {\bibfnamefont {G.}~\bibnamefont {Compagno}},\ }\href
  {https://doi.org/10.1103/PhysRevA.100.012308} {\bibfield  {journal} {\bibinfo
   {journal} {Physical Review A}\ }\textbf {\bibinfo {volume} {100}},\ \bibinfo
  {pages} {012308} (\bibinfo {year} {2019})},\ \Eprint
  {https://arxiv.org/abs/1903.10582} {arXiv:1903.10582} \BibitemShut {NoStop}%
\bibitem [{\citenamefont {Korzekwa}\ \emph {et~al.}(2016)\citenamefont
  {Korzekwa}, \citenamefont {Lostaglio}, \citenamefont {Oppenheim},\ and\
  \citenamefont {Jennings}}]{Korzekwa2016}%
  \BibitemOpen
  \bibfield  {author} {\bibinfo {author} {\bibfnamefont {K.}~\bibnamefont
  {Korzekwa}}, \bibinfo {author} {\bibfnamefont {M.}~\bibnamefont {Lostaglio}},
  \bibinfo {author} {\bibfnamefont {J.}~\bibnamefont {Oppenheim}},\ and\
  \bibinfo {author} {\bibfnamefont {D.}~\bibnamefont {Jennings}},\ }\href
  {https://doi.org/10.1088/1367-2630/18/2/023045} {\bibfield  {journal}
  {\bibinfo  {journal} {New Journal of Physics}\ }\textbf {\bibinfo {volume}
  {18}},\ \bibinfo {pages} {023045} (\bibinfo {year} {2016})},\ \Eprint
  {https://arxiv.org/abs/1506.07875} {arXiv:1506.07875} \BibitemShut {NoStop}%
\bibitem [{\citenamefont {Santos}\ \emph {et~al.}(2019)\citenamefont {Santos},
  \citenamefont {C{\'{e}}leri}, \citenamefont {Landi},\ and\ \citenamefont
  {Paternostro}}]{Santos2019}%
  \BibitemOpen
  \bibfield  {author} {\bibinfo {author} {\bibfnamefont {J.~P.}\ \bibnamefont
  {Santos}}, \bibinfo {author} {\bibfnamefont {L.~C.}\ \bibnamefont
  {C{\'{e}}leri}}, \bibinfo {author} {\bibfnamefont {G.~T.}\ \bibnamefont
  {Landi}},\ and\ \bibinfo {author} {\bibfnamefont {M.}~\bibnamefont
  {Paternostro}},\ }\href {https://doi.org/10.1038/s41534-019-0138-y}
  {\bibfield  {journal} {\bibinfo  {journal} {npj Quantum Information}\
  }\textbf {\bibinfo {volume} {5}},\ \bibinfo {pages} {23} (\bibinfo {year}
  {2019})},\ \Eprint {https://arxiv.org/abs/1707.08946} {arXiv:1707.08946}
  \BibitemShut {NoStop}%
\bibitem [{\citenamefont {Winter}\ and\ \citenamefont
  {Yang}(2016)}]{Winter2016}%
  \BibitemOpen
  \bibfield  {author} {\bibinfo {author} {\bibfnamefont {A.}~\bibnamefont
  {Winter}}\ and\ \bibinfo {author} {\bibfnamefont {D.}~\bibnamefont {Yang}},\
  }\href {https://doi.org/10.1103/PhysRevLett.116.120404} {\bibfield  {journal}
  {\bibinfo  {journal} {Physical Review Letters}\ }\textbf {\bibinfo {volume}
  {116}},\ \bibinfo {pages} {120404} (\bibinfo {year} {2016})},\ \Eprint
  {https://arxiv.org/abs/1506.07975} {arXiv:1506.07975} \BibitemShut {NoStop}%
\bibitem [{\citenamefont {Yuan}\ \emph {et~al.}(2015)\citenamefont {Yuan},
  \citenamefont {Zhou}, \citenamefont {Cao},\ and\ \citenamefont
  {Ma}}]{Yuan2015}%
  \BibitemOpen
  \bibfield  {author} {\bibinfo {author} {\bibfnamefont {X.}~\bibnamefont
  {Yuan}}, \bibinfo {author} {\bibfnamefont {H.}~\bibnamefont {Zhou}}, \bibinfo
  {author} {\bibfnamefont {Z.}~\bibnamefont {Cao}},\ and\ \bibinfo {author}
  {\bibfnamefont {X.}~\bibnamefont {Ma}},\ }\href
  {https://doi.org/10.1103/PhysRevA.92.022124} {\bibfield  {journal} {\bibinfo
  {journal} {Physical Review A}\ }\textbf {\bibinfo {volume} {92}},\ \bibinfo
  {pages} {022124} (\bibinfo {year} {2015})},\ \Eprint
  {https://arxiv.org/abs/1505.04032} {arXiv:1505.04032} \BibitemShut {NoStop}%
\bibitem [{\citenamefont {Sperling}\ and\ \citenamefont
  {Vogel}(2015)}]{Sperling2015}%
  \BibitemOpen
  \bibfield  {author} {\bibinfo {author} {\bibfnamefont {J.}~\bibnamefont
  {Sperling}}\ and\ \bibinfo {author} {\bibfnamefont {W.}~\bibnamefont
  {Vogel}},\ }\href {https://doi.org/10.1088/0031-8949/90/7/074024} {\bibfield
  {journal} {\bibinfo  {journal} {Physica Scripta}\ }\textbf {\bibinfo {volume}
  {90}},\ \bibinfo {pages} {074024} (\bibinfo {year} {2015})},\ \Eprint
  {https://arxiv.org/abs/1004.1944} {arXiv:1004.1944} \BibitemShut {NoStop}%
\bibitem [{\citenamefont {Lambert}\ \emph {et~al.}(2013)\citenamefont
  {Lambert}, \citenamefont {Chen}, \citenamefont {Cheng}, \citenamefont {Li},
  \citenamefont {Chen},\ and\ \citenamefont {Nori}}]{Lambert2013}%
  \BibitemOpen
  \bibfield  {author} {\bibinfo {author} {\bibfnamefont {N.}~\bibnamefont
  {Lambert}}, \bibinfo {author} {\bibfnamefont {Y.-N.}\ \bibnamefont {Chen}},
  \bibinfo {author} {\bibfnamefont {Y.-C.}\ \bibnamefont {Cheng}}, \bibinfo
  {author} {\bibfnamefont {C.-M.}\ \bibnamefont {Li}}, \bibinfo {author}
  {\bibfnamefont {G.-Y.}\ \bibnamefont {Chen}},\ and\ \bibinfo {author}
  {\bibfnamefont {F.}~\bibnamefont {Nori}},\ }\href
  {https://doi.org/10.1038/nphys2474} {\bibfield  {journal} {\bibinfo
  {journal} {Nature Physics}\ }\textbf {\bibinfo {volume} {9}},\ \bibinfo
  {pages} {10} (\bibinfo {year} {2013})}\BibitemShut {NoStop}%
\bibitem [{\citenamefont {Cao}\ \emph {et~al.}(2020)\citenamefont {Cao},
  \citenamefont {Cogdell}, \citenamefont {Coker}, \citenamefont {Duan},
  \citenamefont {Hauer}, \citenamefont {Kleinekath{\"{o}}fer}, \citenamefont
  {Jansen}, \citenamefont {Man{\v{c}}al}, \citenamefont {Miller}, \citenamefont
  {Ogilvie}, \citenamefont {Prokhorenko}, \citenamefont {Renger}, \citenamefont
  {Tan}, \citenamefont {Tempelaar}, \citenamefont {Thorwart}, \citenamefont
  {Thyrhaug}, \citenamefont {Westenhoff},\ and\ \citenamefont
  {Zigmantas}}]{Cao2020}%
  \BibitemOpen
  \bibfield  {author} {\bibinfo {author} {\bibfnamefont {J.}~\bibnamefont
  {Cao}}, \bibinfo {author} {\bibfnamefont {R.~J.}\ \bibnamefont {Cogdell}},
  \bibinfo {author} {\bibfnamefont {D.~F.}\ \bibnamefont {Coker}}, \bibinfo
  {author} {\bibfnamefont {H.-G.}\ \bibnamefont {Duan}}, \bibinfo {author}
  {\bibfnamefont {J.}~\bibnamefont {Hauer}}, \bibinfo {author} {\bibfnamefont
  {U.}~\bibnamefont {Kleinekath{\"{o}}fer}}, \bibinfo {author} {\bibfnamefont
  {T.~L.~C.}\ \bibnamefont {Jansen}}, \bibinfo {author} {\bibfnamefont
  {T.}~\bibnamefont {Man{\v{c}}al}}, \bibinfo {author} {\bibfnamefont
  {R.~J.~D.}\ \bibnamefont {Miller}}, \bibinfo {author} {\bibfnamefont {J.~P.}\
  \bibnamefont {Ogilvie}}, \bibinfo {author} {\bibfnamefont {V.~I.}\
  \bibnamefont {Prokhorenko}}, \bibinfo {author} {\bibfnamefont
  {T.}~\bibnamefont {Renger}}, \bibinfo {author} {\bibfnamefont {H.-S.}\
  \bibnamefont {Tan}}, \bibinfo {author} {\bibfnamefont {R.}~\bibnamefont
  {Tempelaar}}, \bibinfo {author} {\bibfnamefont {M.}~\bibnamefont {Thorwart}},
  \bibinfo {author} {\bibfnamefont {E.}~\bibnamefont {Thyrhaug}}, \bibinfo
  {author} {\bibfnamefont {S.}~\bibnamefont {Westenhoff}},\ and\ \bibinfo
  {author} {\bibfnamefont {D.}~\bibnamefont {Zigmantas}},\ }\href
  {https://doi.org/10.1126/sciadv.aaz4888} {\bibfield  {journal} {\bibinfo
  {journal} {Science Advances}\ }\textbf {\bibinfo {volume} {6}},\ \bibinfo
  {pages} {eaaz4888} (\bibinfo {year} {2020})}\BibitemShut {NoStop}%
\bibitem [{\citenamefont {von Prillwitz}\ \emph {et~al.}(2015)\citenamefont
  {von Prillwitz}, \citenamefont {Rudnicki},\ and\ \citenamefont
  {Mintert}}]{VonPrillwitz2015}%
  \BibitemOpen
  \bibfield  {author} {\bibinfo {author} {\bibfnamefont {K.}~\bibnamefont {von
  Prillwitz}}, \bibinfo {author} {\bibfnamefont {{\L}.}~\bibnamefont
  {Rudnicki}},\ and\ \bibinfo {author} {\bibfnamefont {F.}~\bibnamefont
  {Mintert}},\ }\href {https://doi.org/10.1103/PhysRevA.92.052114} {\bibfield
  {journal} {\bibinfo  {journal} {Physical Review A}\ }\textbf {\bibinfo
  {volume} {92}},\ \bibinfo {pages} {052114} (\bibinfo {year} {2015})},\
  \Eprint {https://arxiv.org/abs/1409.1814} {arXiv:1409.1814} \BibitemShut
  {NoStop}%
\bibitem [{\citenamefont {Dive}\ \emph {et~al.}(2020)\citenamefont {Dive},
  \citenamefont {Koukoulekidis}, \citenamefont {Mousafeiris},\ and\
  \citenamefont {Mintert}}]{Dive2020}%
  \BibitemOpen
  \bibfield  {author} {\bibinfo {author} {\bibfnamefont {B.}~\bibnamefont
  {Dive}}, \bibinfo {author} {\bibfnamefont {N.}~\bibnamefont {Koukoulekidis}},
  \bibinfo {author} {\bibfnamefont {S.}~\bibnamefont {Mousafeiris}},\ and\
  \bibinfo {author} {\bibfnamefont {F.}~\bibnamefont {Mintert}},\ }\href
  {https://doi.org/10.1103/PhysRevResearch.2.013220} {\bibfield  {journal}
  {\bibinfo  {journal} {Physical Review Research}\ }\textbf {\bibinfo {volume}
  {2}},\ \bibinfo {pages} {013220} (\bibinfo {year} {2020})},\ \Eprint
  {https://arxiv.org/abs/1901.08599} {arXiv:1901.08599} \BibitemShut {NoStop}%
\bibitem [{\citenamefont {Blatt}\ and\ \citenamefont {Roos}(2012)}]{Blatt2012}%
  \BibitemOpen
  \bibfield  {author} {\bibinfo {author} {\bibfnamefont {R.}~\bibnamefont
  {Blatt}}\ and\ \bibinfo {author} {\bibfnamefont {C.~F.}\ \bibnamefont
  {Roos}},\ }\href {https://doi.org/10.1038/NPHYS2252} {\bibfield  {journal}
  {\bibinfo  {journal} {Nature Physics}\ }\textbf {\bibinfo {volume} {8}},\
  \bibinfo {pages} {277} (\bibinfo {year} {2012})},\ \Eprint
  {https://arxiv.org/abs/0905.0118} {arXiv:0905.0118} \BibitemShut {NoStop}%
\bibitem [{\citenamefont {Lanyon}\ \emph {et~al.}(2011)\citenamefont {Lanyon},
  \citenamefont {Hempel}, \citenamefont {Nigg}, \citenamefont {Muller},
  \citenamefont {Gerritsma}, \citenamefont {Zahringer}, \citenamefont
  {Schindler}, \citenamefont {Barreiro}, \citenamefont {Rambach}, \citenamefont
  {Kirchmair}, \citenamefont {Hennrich}, \citenamefont {Zoller}, \citenamefont
  {Blatt},\ and\ \citenamefont {Roos}}]{Lanyon2011}%
  \BibitemOpen
  \bibfield  {author} {\bibinfo {author} {\bibfnamefont {B.~P.}\ \bibnamefont
  {Lanyon}}, \bibinfo {author} {\bibfnamefont {C.}~\bibnamefont {Hempel}},
  \bibinfo {author} {\bibfnamefont {D.}~\bibnamefont {Nigg}}, \bibinfo {author}
  {\bibfnamefont {M.}~\bibnamefont {Muller}}, \bibinfo {author} {\bibfnamefont
  {R.}~\bibnamefont {Gerritsma}}, \bibinfo {author} {\bibfnamefont
  {F.}~\bibnamefont {Zahringer}}, \bibinfo {author} {\bibfnamefont
  {P.}~\bibnamefont {Schindler}}, \bibinfo {author} {\bibfnamefont {J.~T.}\
  \bibnamefont {Barreiro}}, \bibinfo {author} {\bibfnamefont {M.}~\bibnamefont
  {Rambach}}, \bibinfo {author} {\bibfnamefont {G.}~\bibnamefont {Kirchmair}},
  \bibinfo {author} {\bibfnamefont {M.}~\bibnamefont {Hennrich}}, \bibinfo
  {author} {\bibfnamefont {P.}~\bibnamefont {Zoller}}, \bibinfo {author}
  {\bibfnamefont {R.}~\bibnamefont {Blatt}},\ and\ \bibinfo {author}
  {\bibfnamefont {C.~F.}\ \bibnamefont {Roos}},\ }\href
  {https://doi.org/10.1126/science.1208001} {\bibfield  {journal} {\bibinfo
  {journal} {Science}\ }\textbf {\bibinfo {volume} {334}},\ \bibinfo {pages}
  {57} (\bibinfo {year} {2011})}\BibitemShut {NoStop}%
\bibitem [{\citenamefont {Schmidt}\ \emph {et~al.}(2005)\citenamefont
  {Schmidt}, \citenamefont {Rosenband}, \citenamefont {Langer}, \citenamefont
  {Itano}, \citenamefont {Bergquist},\ and\ \citenamefont
  {Wineland}}]{Schmidt2005}%
  \BibitemOpen
  \bibfield  {author} {\bibinfo {author} {\bibfnamefont {P.~O.}\ \bibnamefont
  {Schmidt}}, \bibinfo {author} {\bibfnamefont {T.}~\bibnamefont {Rosenband}},
  \bibinfo {author} {\bibfnamefont {C.}~\bibnamefont {Langer}}, \bibinfo
  {author} {\bibfnamefont {W.~M.}\ \bibnamefont {Itano}}, \bibinfo {author}
  {\bibfnamefont {J.~C.}\ \bibnamefont {Bergquist}},\ and\ \bibinfo {author}
  {\bibfnamefont {D.~J.}\ \bibnamefont {Wineland}},\ }\href
  {https://doi.org/10.1126/science.1114375} {\bibfield  {journal} {\bibinfo
  {journal} {Science}\ }\textbf {\bibinfo {volume} {309}},\ \bibinfo {pages}
  {749} (\bibinfo {year} {2005})}\BibitemShut {NoStop}%
\bibitem [{\citenamefont {Rosenband}\ \emph {et~al.}(2007)\citenamefont
  {Rosenband}, \citenamefont {Schmidt}, \citenamefont {Hume}, \citenamefont
  {Itano}, \citenamefont {Fortier}, \citenamefont {Stalnaker}, \citenamefont
  {Kim}, \citenamefont {Diddams}, \citenamefont {Koelemeij}, \citenamefont
  {Bergquist},\ and\ \citenamefont {Wineland}}]{Rosenband2007}%
  \BibitemOpen
  \bibfield  {author} {\bibinfo {author} {\bibfnamefont {T.}~\bibnamefont
  {Rosenband}}, \bibinfo {author} {\bibfnamefont {P.}~\bibnamefont {Schmidt}},
  \bibinfo {author} {\bibfnamefont {D.}~\bibnamefont {Hume}}, \bibinfo {author}
  {\bibfnamefont {W.}~\bibnamefont {Itano}}, \bibinfo {author} {\bibfnamefont
  {T.}~\bibnamefont {Fortier}}, \bibinfo {author} {\bibfnamefont
  {J.}~\bibnamefont {Stalnaker}}, \bibinfo {author} {\bibfnamefont
  {K.}~\bibnamefont {Kim}}, \bibinfo {author} {\bibfnamefont {S.}~\bibnamefont
  {Diddams}}, \bibinfo {author} {\bibfnamefont {J.}~\bibnamefont {Koelemeij}},
  \bibinfo {author} {\bibfnamefont {J.}~\bibnamefont {Bergquist}},\ and\
  \bibinfo {author} {\bibfnamefont {D.}~\bibnamefont {Wineland}},\ }\href
  {https://doi.org/10.1103/PhysRevLett.98.220801} {\bibfield  {journal}
  {\bibinfo  {journal} {Physical Review Letters}\ }\textbf {\bibinfo {volume}
  {98}},\ \bibinfo {pages} {220801} (\bibinfo {year} {2007})}\BibitemShut
  {NoStop}%
\bibitem [{\citenamefont {Monroe}\ \emph {et~al.}(2014)\citenamefont {Monroe},
  \citenamefont {Raussendorf}, \citenamefont {Ruthven}, \citenamefont {Brown},
  \citenamefont {Maunz}, \citenamefont {Duan},\ and\ \citenamefont
  {Kim}}]{Monroe2014}%
  \BibitemOpen
  \bibfield  {author} {\bibinfo {author} {\bibfnamefont {C.}~\bibnamefont
  {Monroe}}, \bibinfo {author} {\bibfnamefont {R.}~\bibnamefont {Raussendorf}},
  \bibinfo {author} {\bibfnamefont {A.}~\bibnamefont {Ruthven}}, \bibinfo
  {author} {\bibfnamefont {K.~R.}\ \bibnamefont {Brown}}, \bibinfo {author}
  {\bibfnamefont {P.}~\bibnamefont {Maunz}}, \bibinfo {author} {\bibfnamefont
  {L.-M.}\ \bibnamefont {Duan}},\ and\ \bibinfo {author} {\bibfnamefont
  {J.}~\bibnamefont {Kim}},\ }\href
  {https://doi.org/10.1103/PhysRevA.89.022317} {\bibfield  {journal} {\bibinfo
  {journal} {Physical Review A}\ }\textbf {\bibinfo {volume} {89}},\ \bibinfo
  {pages} {022317} (\bibinfo {year} {2014})}\BibitemShut {NoStop}%
\bibitem [{\citenamefont {Bruzewicz}\ \emph {et~al.}(2019)\citenamefont
  {Bruzewicz}, \citenamefont {Chiaverini}, \citenamefont {McConnell},\ and\
  \citenamefont {Sage}}]{Bruzewicz2019}%
  \BibitemOpen
  \bibfield  {author} {\bibinfo {author} {\bibfnamefont {C.~D.}\ \bibnamefont
  {Bruzewicz}}, \bibinfo {author} {\bibfnamefont {J.}~\bibnamefont
  {Chiaverini}}, \bibinfo {author} {\bibfnamefont {R.}~\bibnamefont
  {McConnell}},\ and\ \bibinfo {author} {\bibfnamefont {J.~M.}\ \bibnamefont
  {Sage}},\ }\href {https://doi.org/10.1063/1.5088164} {\bibfield  {journal}
  {\bibinfo  {journal} {Applied Physics Reviews}\ }\textbf {\bibinfo {volume}
  {6}},\ \bibinfo {pages} {021314} (\bibinfo {year} {2019})},\ \Eprint
  {https://arxiv.org/abs/1904.04178} {arXiv:1904.04178} \BibitemShut {NoStop}%
\bibitem [{\citenamefont {Ballance}\ \emph {et~al.}(2016)\citenamefont
  {Ballance}, \citenamefont {Harty}, \citenamefont {Linke}, \citenamefont
  {Sepiol},\ and\ \citenamefont {Lucas}}]{Ballance2016}%
  \BibitemOpen
  \bibfield  {author} {\bibinfo {author} {\bibfnamefont {C.~J.}\ \bibnamefont
  {Ballance}}, \bibinfo {author} {\bibfnamefont {T.~P.}\ \bibnamefont {Harty}},
  \bibinfo {author} {\bibfnamefont {N.~M.}\ \bibnamefont {Linke}}, \bibinfo
  {author} {\bibfnamefont {M.~A.}\ \bibnamefont {Sepiol}},\ and\ \bibinfo
  {author} {\bibfnamefont {D.~M.}\ \bibnamefont {Lucas}},\ }\href
  {https://doi.org/10.1103/PhysRevLett.117.060504} {\bibfield  {journal}
  {\bibinfo  {journal} {Physical Review Letters}\ }\textbf {\bibinfo {volume}
  {117}},\ \bibinfo {pages} {060504} (\bibinfo {year} {2016})},\ \Eprint
  {https://arxiv.org/abs/1512.04600} {arXiv:1512.04600} \BibitemShut {NoStop}%
\bibitem [{\citenamefont {Gaebler}\ \emph {et~al.}(2016)\citenamefont
  {Gaebler}, \citenamefont {Tan}, \citenamefont {Lin}, \citenamefont {Wan},
  \citenamefont {Bowler}, \citenamefont {Keith}, \citenamefont {Glancy},
  \citenamefont {Coakley}, \citenamefont {Knill}, \citenamefont {Leibfried},\
  and\ \citenamefont {Wineland}}]{Gaebler2016}%
  \BibitemOpen
  \bibfield  {author} {\bibinfo {author} {\bibfnamefont {J.~P.}\ \bibnamefont
  {Gaebler}}, \bibinfo {author} {\bibfnamefont {T.~R.}\ \bibnamefont {Tan}},
  \bibinfo {author} {\bibfnamefont {Y.}~\bibnamefont {Lin}}, \bibinfo {author}
  {\bibfnamefont {Y.}~\bibnamefont {Wan}}, \bibinfo {author} {\bibfnamefont
  {R.}~\bibnamefont {Bowler}}, \bibinfo {author} {\bibfnamefont {A.~C.}\
  \bibnamefont {Keith}}, \bibinfo {author} {\bibfnamefont {S.}~\bibnamefont
  {Glancy}}, \bibinfo {author} {\bibfnamefont {K.}~\bibnamefont {Coakley}},
  \bibinfo {author} {\bibfnamefont {E.}~\bibnamefont {Knill}}, \bibinfo
  {author} {\bibfnamefont {D.}~\bibnamefont {Leibfried}},\ and\ \bibinfo
  {author} {\bibfnamefont {D.~J.}\ \bibnamefont {Wineland}},\ }\href
  {https://doi.org/10.1103/PhysRevLett.117.060505} {\bibfield  {journal}
  {\bibinfo  {journal} {Physical Review Letters}\ }\textbf {\bibinfo {volume}
  {117}},\ \bibinfo {pages} {060505} (\bibinfo {year} {2016})},\ \Eprint
  {https://arxiv.org/abs/1604.00032} {arXiv:1604.00032} \BibitemShut {NoStop}%
\bibitem [{\citenamefont {Leibfried}\ \emph {et~al.}(1996)\citenamefont
  {Leibfried}, \citenamefont {Meekhof}, \citenamefont {King}, \citenamefont
  {Monroe}, \citenamefont {Itano},\ and\ \citenamefont
  {Wineland}}]{Leibfried1996}%
  \BibitemOpen
  \bibfield  {author} {\bibinfo {author} {\bibfnamefont {D.}~\bibnamefont
  {Leibfried}}, \bibinfo {author} {\bibfnamefont {D.~M.}\ \bibnamefont
  {Meekhof}}, \bibinfo {author} {\bibfnamefont {B.~E.}\ \bibnamefont {King}},
  \bibinfo {author} {\bibfnamefont {C.}~\bibnamefont {Monroe}}, \bibinfo
  {author} {\bibfnamefont {W.~M.}\ \bibnamefont {Itano}},\ and\ \bibinfo
  {author} {\bibfnamefont {D.~J.}\ \bibnamefont {Wineland}},\ }\href
  {https://doi.org/10.1103/PhysRevLett.77.4281} {\bibfield  {journal} {\bibinfo
   {journal} {Phys. Rev. Lett.}\ }\textbf {\bibinfo {volume} {77}},\ \bibinfo
  {pages} {4281} (\bibinfo {year} {1996})}\BibitemShut {NoStop}%
\bibitem [{\citenamefont {Gardiner}\ \emph {et~al.}(1997)\citenamefont
  {Gardiner}, \citenamefont {Cirac},\ and\ \citenamefont
  {Zoller}}]{Gardiner1997}%
  \BibitemOpen
  \bibfield  {author} {\bibinfo {author} {\bibfnamefont {S.~A.}\ \bibnamefont
  {Gardiner}}, \bibinfo {author} {\bibfnamefont {J.~I.}\ \bibnamefont
  {Cirac}},\ and\ \bibinfo {author} {\bibfnamefont {P.}~\bibnamefont
  {Zoller}},\ }\href {https://doi.org/10.1103/PhysRevA.55.1683} {\bibfield
  {journal} {\bibinfo  {journal} {Physical Review A}\ }\textbf {\bibinfo
  {volume} {55}},\ \bibinfo {pages} {1683} (\bibinfo {year}
  {1997})}\BibitemShut {NoStop}%
\bibitem [{\citenamefont {Ben-Kish}\ \emph {et~al.}(2003)\citenamefont
  {Ben-Kish}, \citenamefont {DeMarco}, \citenamefont {Meyer}, \citenamefont
  {Rowe}, \citenamefont {Britton}, \citenamefont {Itano}, \citenamefont
  {Jelenkovi{\'{c}}}, \citenamefont {Langer}, \citenamefont {Leibfried},
  \citenamefont {Rosenband},\ and\ \citenamefont {Wineland}}]{Ben-Kish2003}%
  \BibitemOpen
  \bibfield  {author} {\bibinfo {author} {\bibfnamefont {A.}~\bibnamefont
  {Ben-Kish}}, \bibinfo {author} {\bibfnamefont {B.}~\bibnamefont {DeMarco}},
  \bibinfo {author} {\bibfnamefont {V.}~\bibnamefont {Meyer}}, \bibinfo
  {author} {\bibfnamefont {M.}~\bibnamefont {Rowe}}, \bibinfo {author}
  {\bibfnamefont {J.}~\bibnamefont {Britton}}, \bibinfo {author} {\bibfnamefont
  {W.~M.}\ \bibnamefont {Itano}}, \bibinfo {author} {\bibfnamefont {B.~M.}\
  \bibnamefont {Jelenkovi{\'{c}}}}, \bibinfo {author} {\bibfnamefont
  {C.}~\bibnamefont {Langer}}, \bibinfo {author} {\bibfnamefont
  {D.}~\bibnamefont {Leibfried}}, \bibinfo {author} {\bibfnamefont
  {T.}~\bibnamefont {Rosenband}},\ and\ \bibinfo {author} {\bibfnamefont
  {D.~J.}\ \bibnamefont {Wineland}},\ }\href
  {https://doi.org/10.1103/PhysRevLett.90.037902} {\bibfield  {journal}
  {\bibinfo  {journal} {Physical Review Letters}\ }\textbf {\bibinfo {volume}
  {90}},\ \bibinfo {pages} {037902} (\bibinfo {year} {2003})}\BibitemShut
  {NoStop}%
\bibitem [{\citenamefont {Szalay}(2015)}]{Szalay2015}%
  \BibitemOpen
  \bibfield  {author} {\bibinfo {author} {\bibfnamefont {S.}~\bibnamefont
  {Szalay}},\ }\href {https://doi.org/10.1103/PhysRevA.92.042329} {\bibfield
  {journal} {\bibinfo  {journal} {Phys. Rev. A}\ }\textbf {\bibinfo {volume}
  {92}},\ \bibinfo {pages} {042329} (\bibinfo {year} {2015})}\BibitemShut
  {NoStop}%
\bibitem [{Note1()}]{Note1}%
  \BibitemOpen
  \bibinfo {note} {See Supplemental Information at \protect \href
  {https://github.com/ImperialCQD/Certifying-Multilevel-Coherence-in-the-Motional-State-of-a-Trapped-Ion}{https://github.com/ImperialCQD/Certifying-Multilevel-Coherence-in-the-Motional-State-of-a-Trapped-Ion}}\BibitemShut
  {NoStop}%
\bibitem [{\citenamefont {Casella}\ and\ \citenamefont
  {Berger}(2002)}]{Casella2002}%
  \BibitemOpen
  \bibfield  {author} {\bibinfo {author} {\bibfnamefont {G.}~\bibnamefont
  {Casella}}\ and\ \bibinfo {author} {\bibfnamefont {R.~L.}\ \bibnamefont
  {Berger}},\ }\href@noop {} {\emph {\bibinfo {title} {{Statistical
  inference}}}},\ \bibinfo {edition} {2nd}\ ed.\ (\bibinfo  {publisher}
  {Thomson Learning},\ \bibinfo {year} {2002})\BibitemShut {NoStop}%
\bibitem [{\citenamefont {Chan}\ \emph {et~al.}(2020)\citenamefont {Chan},
  \citenamefont {Chen}, \citenamefont {Li}, \citenamefont {Wong},\ and\
  \citenamefont {Yau}}]{Chan2020}%
  \BibitemOpen
  \bibfield  {author} {\bibinfo {author} {\bibfnamefont {L.~H.}\ \bibnamefont
  {Chan}}, \bibinfo {author} {\bibfnamefont {K.}~\bibnamefont {Chen}}, \bibinfo
  {author} {\bibfnamefont {C.}~\bibnamefont {Li}}, \bibinfo {author}
  {\bibfnamefont {C.~W.}\ \bibnamefont {Wong}},\ and\ \bibinfo {author}
  {\bibfnamefont {C.~Y.}\ \bibnamefont {Yau}},\ }\href
  {https://doi.org/10.1080/00949655.2019.1700987} {\bibfield  {journal}
  {\bibinfo  {journal} {Journal of Statistical Computation and Simulation}\
  }\textbf {\bibinfo {volume} {90}},\ \bibinfo {pages} {747} (\bibinfo {year}
  {2020})}\BibitemShut {NoStop}%
\end{thebibliography}
\end{document}